\documentstyle[12pt,aps,pictex]{revtex}  

\begin{document}

\draft

\title{Quantum scalar field in FRW Universe with constant electromagnetic 
background}

\author{S.P. Gavrilov\thanks{On leave from Tomsk Pedagogical
University, 634041 Tomsk, Russia; present  address: Dept. Fisica, CCE,
Universidade Estadual de Londrina, CP 6001. CEP 86051-990, Londrina, PR,
Brasil;
e-mail: 
gavrilov@fisica.uel.br}, D.M. Gitman\thanks{e-mail: 
gitman@fma.if.usp.br} and S.D. Odintsov
\thanks{On leave from  Tomsk Pedagogical
University, 634041 Tomsk, Russia; present  address: Dept. de Fisica,
Universidad del Valle, AA 25360 Cali, Columbia;
e-mail: 
odintsov@quantum.univalle.edu.co}} 

\address{Instituto de F\'{\i}sica, Universidade de S\~ao Paulo, \\ 
Caixa Postal 66318, 05315-970-S\~ao Paulo, S.P. Brasil
}

\date{\today}

\maketitle

\begin{abstract}
We discuss massive scalar field with conformal coupling in
Friedmann-Robertson-Walker (FRW)  Universe of special type with
constant electromagnetic field. Treating an external
gravitational-electromagnetic background exactly, at first time the
proper-time representations for out-in, in-in, and out-out scalar  Green
functions  are explicitly constructed as proper-time integrals over
the corresponding (complex) contours. The vacuum-to-vacuum transition
amplitudes and number of created  particles are found and vacuum
instability is discussed. The mean values of the current and
energy-momentum tensor  are evaluated, and different
approximations for them are investigated. The back reaction of the
particles created to the electromagnetic field is estimated in
different regimes. The connection between proper-time method and
effective action is outlined. The effective action in scalar QED in
weakly-curved FRW Universe (De Sitter space) with weak constant
electromagnetic field is found as derivative expansion over curvature
and electromagnetic field strength. Possible further applications of
the results are briefly mentioned. 
\end{abstract}

\section{Introduction}
It is quite well known fact that quantum field theory in an external
background is, generally speaking, theory with unstable vacuum. The
vacuum instability leads to many interesting features, among which
particles creation from vacuum is one of the
most beautiful non-perturbative phenomena in  quantum field
theory. Furthermore, in interacting theories the vacuum instability
may lead to quantum processes which are prohibited if the vacuum is
stable. One ought to say that all the above mentioned peculiarities
can not be reveal in  frames of the perturbation theory with regards
to the external background, one has to
treat it exactly. The latter has been realized
long ago by Schwinger \cite{Sch1} on the example of quantum
electrodynamics in the constant electric field. The particles creation
in this case has been calculated explicitly.

In  quantum field theory with unstable vacuum it
is necessary to construct different kinds of Green functions (GF), e.g. besides 
the
causal GF (out-in GF) one has to use so called in-in GF, out-out
GF, and so on \cite{BdW,Git1,FG} (for a review and technical details see
\cite{FGS}). General methods of such GF construction in electromagnetic
(EM) background have been  developed in \cite{Git1,FG}. The possible 
generalization of the
formalism to an external gravitational
background has been given in ref. \cite{BFG}. Since the paper
\cite{Sch1} it is known that causal (out-in) GF may be
presented as a proper-time integral over a real infinite
contour. At the same time, in the instable vacuum  the in-in
and out-out GF differ from the causal one. One can show
\cite{GGS} that these functions may be presented by the same
proper-time integral (with the same integrand) but over another
contours in the complex proper-time plane. The complete set of GF
mentioned is necessary for the construction of Furry picture in interacting 
theories,
and even in non-interacting cases one has to use them to define, for
example, the back reaction of particles created and to construct
different kinds of effective actions.  

It may be likely that early Universe (EU) is filled with some type of
electromagnetic fields. For example, recently (see \cite{TW,GGV} and 
references therein) the possibility of existence and role of
primordial magnetic fields in EU have been discussed. From
another point the possibility of existence of electromagnetic field in
the EU has been discussed long ago in \cite{SD,BO1}. It
has been shown there that the presence of the electrical field in the EU 
significally increases the gravitational particle creation from the
vacuum. In principle, this process may be considered as a source for
the dominant part of the Universe mass. 

Having in mind the above cosmological motivations it is getting
interesting to study the quantum field theory in curved background
with electromagnetic field (of special form to be able to solve the
problem analytically). In the present paper we are going to consider a
massive charged scalar field with conformal coupling in the expanding FRW
Universe with the scale factor $\Omega (\eta)$ (in terms of the
conformal time) $\Omega^2 (\eta)=b^2\eta^2+a^2$. Such a scale factor
corresponds to the expanding radiation dominated FRW Universe. In
terms of physical time $t$ the corresponding metric may be written as
following:
\begin{equation}\label{1}
ds^2=dt^2-\Omega^2(t)(dx^2+dy^2+dz^2)\;,
\end{equation}
where for small times $|t|\ll a^2/b,\; \Omega^2(t)\simeq
a^2[1+(bt/a^2)^2]$ and for large times $|t|\gg a^2/b,\; \Omega^2(t)\simeq
2b|t|$ (see \cite{SD}). Moreover, such FRW Universe will be filled by
the constant electromagnetic field (the precise form of this field is
given in the next section). Thus, we start from the charged scalar theory in above
background. Making conformal transformation of scalar theory (which
works generally speaking for an arbitrary dimensionality of spacetime
if correspondent conformal coupling is chosen), we remain with the
theory in flat background but with time-dependent mass. For
generality, we  consider such a theory in $d$-dimensions (what may
by useful also for Kaluza-Klein type theories \cite{BOS}). The
embedding of such a theory to $d$-dimensional constant EM
field  will be considered. (Of course, our main physical interest
will be related with  $d=4$ where the Maxwell theory is
conformally invariant, and one can start from the FRW Universe
(\ref{1}) with constant EM field from the very beginning, before
the conformal transformation). In addition, in $d=4$ the conformal
coupling (i.e. the choice of scalar gravitational coupling $\xi=1/6$)
is UV fixed point of renormalization group (RG) \cite{BOS,BO2} for a
variety of interacting theories (for example, for charged
self-interacting scalar theory it is IR stable fixed point of RG
\cite{BO2}). Hence, there appears the additional motivation to start
from conformal coupling $\xi=1/6$, as anyway at high energies (strong
curvatures) \cite{BOS,BO2} $\xi$ tends to this value. Hence, in
Sect. II we solve $d$-dimensional Klein-Gordon equation for the theory in
the constant EM field and with time-dependent mass to get complete sets of
solutions classified as particles and antiparticles at $t\rightarrow
\pm \infty$. Using them we  construct all necessary  
GF for the scalar field  as Schwinger type proper-time
 integrals. All GF have the same integrand and differ
by the contours of integration in the complex proper-time plane.
 As far as we know
 that it is first explicit example for proper-time representation for
complete set of scalar GF in gravitational-electromagnetic background (for
pure electromagnetic background it was calculated in \cite{GGS} and 
 in pure gravitational background, see for example  \cite{BKO}). 
Using the exact solutions obtained we discuss particles
creation under the combined effect of
an external gravitational and electromagnetic fields. By means of  
GF  we discuss the back reaction of the particles created and the
effective action. Besides, in the Sect. IV we apply
renormalization group (RG) method for calculation of the effective
action in constant curvature spacetime filled by the constant EM
field. The typical example of such spacetime is de Sitter Universe
(exponentially expanding one). As specific theory, we consider scalar
quantum electrodynamics (QED) on above-described background. The RG
improved effective action is calculated as the derivative expansion
over the external fields (hence, in this Sect. we consider
weakly-curved spacetime filled by weak EM fields). Such an effective
action represents the extension of the well-known Schwinger effective
Lagrangian for the case of curved spacetime.

In the Conclusion the summary of the results obtained is
presented. Some further prospects and possible applications are also discussed.

\section{Exact solutions and Green functions in the external time-dependent 
background}

In this section we will  present exact solutions and GF   of the charged
scalar field in the external constant uniform electromagnetic background.
In addition, the former field will be considered in the time-dependent
mass-like potential, which effectively reproduces effects of a
gravitational background. 

The scalar field obeys the  Klein-Gordon equation:
\begin{equation}\label{a1}
\left(\hat{P}^2 -M^2 \Omega^2 (\eta)\right) \varphi = 0 
\end{equation}
where $\,\hat{P}_\mu = i\partial_\mu - q\,A_\mu;\;q=-|e|\,$ (for electron);
$\,x^0\equiv\eta,\; \partial_0 \equiv\partial_\eta;\,\varphi(x)\,$ is
the scalar field; the dimensionality of the space is chosen to be equal to
$\,d=D+1,\,d\geq 2\,$ and the Minkowswski tensor has a form
$\eta_{\mu
\nu}={\rm diag}\left(1,-1,-1,\ldots \right).$
The time-dependent potential term is chosen as
$\,\Omega^2(\eta) = b^2\eta^2 +a^2\,$. The external electromagnetic 
field will be chosen as the following:
constant uniform electric field
\begin{equation}\label{a2}
F_{0D} =E \;\left(F_{\mu \nu} = \partial_\mu\,A_{\nu} -
     \partial _{\nu} A_\mu\right)\;,
\end{equation}
and magnetic field ( for $\, d>3\,$), which is defined by $\,[d/2]-1\,$
invariants of the Lorentz transformations. This magnetic field is
given by the corresponding components $\, H_j \,$ (see for details \cite{GG1}):
\begin{equation}\label{a3}
F^{(H)}_{\mu \nu} = \sum_{\jmath=1}^{[d/2]-1} H_\jmath 
   \left(\delta_\mu^{2\jmath}\delta_{\nu}^{2\jmath-1} -
    \delta^{2\jmath}_\nu \delta_\mu^{2\jmath-1}\right)\;.
\end{equation}  
For such a field we select the  following potentials:
\begin{equation}\label{a4}
A_0 = 0 , \; A_D = Ex^0, \; A_i = A_i^{\perp}= -
H_j\,x_{i+1} \,\delta_{i,2j-1} ,\; 
    j=1,\ldots, [d/2]-1,
    \; i=1, \ldots, D-1.
\end{equation}
In this case solutions
$\varphi $ of the equation (\ref{a1}) are related with ones $\varphi '$ 
with some other choice of the
potentials $A'_{\mu}$ for the same electromagnetic field, by the relation   
\begin{equation}\label{a5}
\varphi ' = \exp\left ({iq \int_{x_c}^{x} \left(A_\mu - A_\mu '\right)
    dx^\mu} \right )\,\varphi \;,
\end{equation}
where the integral is taken along the line: $
\partial_\nu \int_{x_c}^{x} \left(A_\mu - A'_\mu\right) 
    dx^\mu = A_\nu - A'_\nu $.

Let us discuss first the physical motivations behind the Eq. (\ref{a1}). For
$\,b^2=0,\; a^2=1\,$ we have $d$-dimensional scalar field in the
constant electromagnetic field. If in this case $\,d>4\,$ we have
typical Kaluza-Klein type theory in external \ EM \ background (for review
of quantum Kaluza-Klein theories, see, for example \cite{BOS}).

The case $\,d=4\,$, and $\,b^2\neq 0\,$ is of special interest for us.
In this case one can consider charged scalar field in the conformally-flat
Universe (with scale factor $\,\Omega (\eta)= \sqrt{b^2\eta^2+a^2})\,$
filled by the constant \ EM \ field. Choosing the scalar-gravitational
coupling constant $\,\xi=1/6\,$ (conformal coupling) and making the
standard conformal transformation of the gravitational metric and
scalar field, we come to the theory in flat spacetime with time-dependent
mass. The corresponding field equation is given by  (\ref{a1}) (EM  field
should not be transformed under the conformal transformation). Note
that $\,\xi=1/6\,$ is $\,UV\,$ fixed point of  RG  for a variety of
theories \cite{BO2,BOS}, so such conformal transformation may be used also
for interacting theories (let us remind that
$\,(\varphi^+\varphi)^2$-term is conformally invariant in the
Lagrangian). Hence, $\,d=4\,$ case in Eq. (\ref{a1}) actually corresponds to
the quantum scalar field in the expanding FRW-Universe with constant \
EM \ field. However, for generality we leave $\,d\,$ to be an arbitrary
integer number.

As was already said,  when  quantum fields are considered in time-dependent 
backgrounds
(electromagnetic or gravitational ones) one has to  construct
 different Green functions. To this end one has to find special sets
of exact solutions of the equation (\ref{a1}). Here we are going to
describe such solutions.  
The functions $\varphi (x)\,$ can be presented in form:
\begin{equation}\label{a6}
\varphi_{p_Dn} (x) = \varphi _{p_Dn} (x_\parallel)\,
     \varphi_n(x_\perp)\quad ,
\end{equation}
where nonzero $\,x_\perp^i=x^i,\quad i=1,\ldots, D-1,\quad x_\parallel^{\mu} = 
x^{\mu},\;
\mu=0,D\,$ and
\begin{equation}\label{a7}
\hat{{\bf P}}_{\perp}^2 \varphi_n(x_\perp) =  {\bf P}^2_{\perp}
     \varphi_n(x_{\perp}),\quad P_{\perp}^i=P^i, \quad
       i=1,\ldots, D-1\;,
\end{equation}
\begin{equation}\label{a8}
\int \varphi^*_n(x_\perp) \varphi_{n'} (x_\perp)
    d{\bf x}_\perp = \delta_{nn'}\;.
\end{equation}
Here $\,n=\left(n_1,\ldots, n_{[d/2]-1}, p_1,p_3,\ldots,
p_{2[(d-1)/2]-1}\right)\,$ is complete set of the quantum numbers in the
space $\,x_\perp\,$. The dimensionality of $n$ is equal to $d-2$. 
The expression $\, {\bf P}\,^2_\perp\,$ is defined as
\begin{eqnarray}\label{a9} 
&& {\bf P}^2_\perp =  \sum_{j=1}^{[d/2]-1}\omega
_j+\omega_0,\;\;\;\omega_0
 =\left\{ \begin{array}{ll}
0, &\mbox{ d is even} \\
p^2_{d-2}, &\mbox{d is odd}\;,
\end{array}
\right. \\
&&\omega_j =\left\{ \begin{array}{ll}
 |qH_j|(2n_j+1),\;n_j=0,1,\ldots\;\;, & H_j\neq 0 \\
 p^2_{2j-1}+p^2_{2j}\;\;,&H_j=0
\end{array}\right.\;,\nonumber
\end{eqnarray}
where
in the presence of the magnetic field some momenta $p_{2j}$ have to be
replaced by the discrete quantum numbers $n_j$. The number of these
momenta $p_{2j}$ corresponds to the number of nonzero parameters
$H_j$.
Note that $\,n\,$ includes the discrete $\,n_j\,$ and
continuous $\,p_{2j-1}\,$:
\begin{eqnarray}\label{a10}
\left(\hat{P}^2_{2j-1} + \hat{P}^2_{2j}\right) \varphi_n(x_\perp) &=& 
   \omega_j\varphi_n(x_\perp)\;,\nonumber \\[0.3cm]
\hat{P}_{2j-1} \,\varphi_n (x_\perp) &=& p_{2j-1}\,\varphi_n (x_\perp)\;.
\end{eqnarray}

Let us write
\begin{equation}\label{a11}
\varphi_{p_D n} (x_\parallel) = \frac{1}{\sqrt{2\pi}}\,
      e^{-ip_D x^D} \varphi_{p_D n} (x^0)\quad ,
\end{equation}
where
\[
 \varphi_{p_D n}(x^0) = \varphi_{p_D n} (x^0 ,
p_z)|_{p_z=0}\;,
\]
and $\varphi(x^0, p_z)$ is a solution of equation 
\begin{equation}\label{a12}
\left[\left(i\frac{\partial}{\partial \widetilde{\eta}}\right)^2 - 
     \left(p_z - \rho \widetilde{\eta}\right)^2
     - \rho\lambda\right] \varphi_{p_D n} 
      \left(x^0, p_z\right)=0\;,
\end{equation}
with
\[
\widetilde{\eta} =  x^0 - \frac{1}{\rho^2} \,qEp_D ,
  \;\rho^2 = (qE)^2 + (bM)^2, \quad \rho\lambda = (aM)^2 +
     p^2_D \,\frac{(bM)^2}{\rho^2} +  {\bf P}^2_{\perp}\; .
\]
Orthonormalized and classified as particles (+) and antiparticles (-)
at $x^0\rightarrow \pm \infty$ solutions of the equation (\ref{a12})
have the form \cite{Nik}
\begin{eqnarray}\label{a13}
&& {^-_+}\varphi_{p_Dn} \left(x^0, p_z\right) = BD_\nu [\pm(1-i)\xi],\;\;
{^+_-}\varphi_{p_Dn}\left(x^0, p_z\right) = B D_{-\nu-1}
   \left[\pm(1+i)\xi\right]\;,\nonumber \\
&& \xi = \frac{1}{\sqrt{\rho}}\,\left(\rho\widetilde{\eta} - 
        p_z\right),\; \nu= \frac{i\lambda}{2} -\frac{1}{2}\;,\quad 
B=(2\rho)^{-1/4}\,
   \exp\left(-\frac{\pi\lambda}{8}\right)\;,
\end{eqnarray}
see \cite{GG1} for additional arguments advocating such a classification.

One can find   decomposition coefficients
$G\left({}_{\zeta}|{}^{\zeta '}\right)$ of the out-solutions in the in-solutions,
\begin{equation}\label{e10}
{}^{\zeta}\varphi(x)={}_{+}\varphi(x)G\left({}_{+}|{}^{\zeta}\right)-{}_{-}\varphi
(x)
G\left({}_{-}|{}^{\zeta}\right)\;,\;\;\zeta=\pm\;.
\end{equation}
The matrices $G\left({}_{\zeta}|{}^{\zeta'}\right)$ obey the
following relations,
\begin{eqnarray}\label{e10a}
&&G\left({}_{\zeta}|{}^{+}\right)G\left({}_{\zeta}|{}^{+}\right)^{\dagger}- 
G\left({}_{\zeta}|{}^{-}\right)G\left({}_{\zeta}|{}^{-}\right)^{\dagger}=
\zeta I ,\nonumber \\                  
&&G\left({}_{+}|{}^{+}\right)G\left({}_{-}|{}^{+}\right)^{\dagger}-
G\left({}_{+}|{}^{-}\right)G\left({}_{-}|{}^{-}\right)^{\dagger}=0\;.
\end{eqnarray}
The latter can be derived from
the orthonormality conditions. One  can easily see
that 
$G\left({}_{\zeta}|{}^{\zeta'}\right)$
are diagonal,
\begin{equation}\label{e37a}
G\left({}_{\zeta}|{}^{\zeta'}\right)_{ll'}=\delta_{l,l'}
\;g\left({}_{\zeta}|{}^{\zeta'}\right)\;, \;\;l=(p_D,n)\;, l'=(p'_D,n')\;,
\end{equation}
where
\begin{equation}\label{a15}
g(_{\zeta}|^{\zeta'}) = {}_{\zeta}\varphi^*_{p_D n} (x^0,p_z) \,
    i \stackrel{\leftrightarrow}{\partial}_0 \, {}^{\zeta'}\varphi_{p_D n} 
     (x^0, p_z)\; .
\end{equation}

As was already said, in case of theories with unstable vacuum one has
to study different types of GF, which we define via
$\Delta,\;\Delta^c,\;\Delta^{\mp},\;\Delta^c_{in},\;\Delta^{\mp}_{in}$,
and so on, according  \cite{Git1,FG,FGS}. For example, the field
theoretical definitions of causal out-in, in-in and out-out GF
have the form
\begin{eqnarray}\label{Gr}
&& \Delta^c (x,x') =c_v^{-1}i<0,out|T\phi (x)\phi
^{\dagger}(x')|0,in>, \;\;c_v=<0,out|0,in>\;,\\
&&  \Delta^c_{in} (x,x') =i<0,in|T\phi (x)\phi
^{\dagger}(x')|0,in>\;,\\
&&  \Delta^c_{out} (x,x') =i<0,out|T\phi (x)\phi
^{\dagger}(x')|0,out>\;,
\end{eqnarray}
where $\phi (x)$ is a quantized scalar field, 
$|0,in>$ and $|0,out>$ are initial and final vacua and $c_v$ is 
vacuum to vacuum transition amplitude. 
In principle, one can calculate all of GF   
using sets of solutions  (\ref{a6}) and some relations
between the functions,
\begin{eqnarray}
&& \Delta^c (x,x') = 
      \theta \left(x_0-{{x'}_0}\right) \,{\Delta^-}(x,x') -
     \theta\left({{x'}_0} - x_0\right){\Delta^+} (x, x')\quad , \label{a16}\\
&& \Delta (x,x') = i\left[\phi (x),\phi^{\dagger}(x')\right]_- =
{\Delta^{-}} (x,x') + {\Delta^+} (x, x')\quad ,\label{a17}\\
&& {\Delta^-} (x, x') = i\int^{+\infty}_{-\infty} dp_D\, \sum_n\, 
   {^+\varphi_{p_D n}} (x) g(_+|^+)^{-1}{}_+\varphi^*_{p_D n}(x')
    \quad , \nonumber\\[0.3cm]
&& {\Delta^+} (x, x') = -i \int^{+\infty}_{-\infty} dp_D\,\sum_n
       {_{-}\varphi}_{p_D n} (x) \left[g(_-|^-)^{-1}\right]^*{}^-\varphi_{p_D n}^*
           (x')\quad ,  \label{a18}
\end{eqnarray}
\begin{eqnarray}\label{a19}
&&
{\Delta^c}_{in} (x,x') = \theta(x_0-{x'}_0) \Delta^-_{in} 
     (x,x') - \theta ({x'}_0 - x_0) {\Delta^+}_{in} (x,x')\quad 
,\nonumber\\[0.3cm]
&&\Delta^{\mp}_{in} (x, x') = \pm i \int^{+\infty}_{-\infty}\,
   dp_D \sum_n \,{}_{\pm}\varphi_{p_D n} (x) \,{_\pm\varphi^*_{p_D n}}
     (x')\quad ,\\
&&
{\Delta^c}_{out} (x,x') = \theta(x_0-{x'}_0) \Delta^-_{out} 
     (x,x') - \theta ({x'}_0 - x_0) {\Delta^+}_{out} (x,x')\quad 
,\nonumber\\[0.3cm]
&&\Delta^{\mp}_{out} (x, x') = \pm i \int^{+\infty}_{-\infty}\,
   dp_D \sum_n \,{}^{\pm}\varphi_{p_D n} (x) \,{^\pm\varphi^*_{p_D n}}
     (x')\quad .  
\end{eqnarray}
Below we are going to do such calculations in the case under
consideration. First, one can remark that the functions $\Delta^\mp$
and $\Delta^\mp_{in}$  can be  presented as follows
\begin{eqnarray}\label{a20}
&&\pm\Delta^\mp (x,x')  =  \Delta^c (x,x') \pm \theta  
     (\mp (x_0-x_0')) \Delta (x,x')\quad ,\\[0,3cm]
&&\pm\Delta^\mp_{in}  (x,x')  =  \Delta^c_{in} (x,x') \pm \theta  
     (\mp (x_0-x_0')) \Delta (x,x')\quad ,\label{a21}\\[0,3cm]
&&\pm\Delta^\mp_{out}  (x,x')  =  \Delta^c_{out} (x,x') \pm \theta  
     (\mp (x_0-x_0')) \Delta (x,x')\quad ,\label{a21a}\\[0,3cm]
&&\Delta^c_{in} (x,x')  =  \Delta^c (x,x')
     -\Delta^{a}(x,x')\quad ,\label{a22}\\[0,3cm]
&&\Delta^c_{out} (x,x')  =  \Delta^c (x,x')
     -\Delta^{p}(x,x')\quad ,\label{a22a}\\[0,3cm]
&& \Delta^a (x,x')  = -i \int^{+\infty}_{-\infty}
    dp_D {\sum_n} {_-\varphi_{p_D n}} (x) \,
      \left[g(_-|^-)^{-1} g(_+|^-)\right]^*
       {_+\varphi^*_{p_D n}} (x')\quad ,\label{a23}\\
&& \Delta^p (x,x')  = -i \int^{+\infty}_{-\infty}
    dp_D {\sum_n} {^+\varphi_{p_D n}} (x) \,
      g(_+|^+)^{-1} g(_+|^-)
       {^-\varphi^*_{p_D n}} (x')\quad .\label{a23a}
\end{eqnarray}
The coefficients (\ref{a15}) do not depend on $p_z$, thus, one can
 present the functions $\Delta^\mp$ and $\Delta^{a,p}$
 in the  convenient form 
\begin{equation}\label{a24}
\Delta^{\mp, a,p} (x,x') = \int^{+\infty}_{-\infty} \,
      dz\, \int^{+\infty}_{-\infty}\,
        \frac{dp_D}{2\pi} \, e^{-ip_D y^D}\Delta_Q^{\mp,a,p}
    \; ,\quad 
     y_\mu = x_\mu - x'_\mu\;,
\end{equation}
\[
\Delta_Q^{\mp,a,p}=\Delta_Q^{\mp,a,p}
            \left(\widetilde{\eta},x_{\perp}, \widetilde{\eta'},
             x'_{\perp}, z-z', p_D\right) \;, 
\]
\begin{eqnarray}\label{a25}
&&
\Delta^-_Q =
     i\sum_n \,\int^{+\infty}_{-\infty} dp_z \,  {^+\varphi_{p_D n}}
       (\widetilde{\eta}, x_{\perp},z, p_z)g(_+|^+)^{-1}{}_+\varphi^*_{p_D n} 
         \left(\widetilde{\eta'} , x'_{\perp}, z', p_z\right)\quad 
,\nonumber\\[0.3cm]
&&\Delta^+_Q =
    - i\sum_n \,\int^{+\infty}_{-\infty} dp_z \,  {_-\varphi_{p_D n}}
       (\widetilde{\eta}, x_{\perp},z, p_z)
\left[g(_-|^-)^{-1}\right]^*
{}^-\varphi^*_{p_D n} 
         \left(\widetilde{\eta'} , x'_{\perp}, z', p_z\right)\quad 
,\nonumber\\[0.3cm]
&&\Delta^a_Q =
    -i\sum_n \,\int^{+\infty}_{-\infty} dp_z \,  {_-\varphi_{p_D n}}
       (\widetilde{\eta}, x_{\perp},z, p_z)
\left[g(_-|^-)^{-1} g(_+|^-)\right]^*
{}_+\varphi^*_{p_D n} 
         \left(\widetilde{\eta'}, {x'}_{\perp},z',
       p_z\right)\quad ,\label{a26}\\
&&\Delta^p_Q =
    -i\sum_n \,\int^{+\infty}_{-\infty} dp_z \,  {^+\varphi_{p_D n}}
       (\widetilde{\eta}, x_{\perp},z, p_z)
g(_+|^+)^{-1} g(_+|^-)
{}^-\varphi^*_{p_D n} 
         \left(\widetilde{\eta'}, {x'}_{\perp},z',
       p_z\right)\quad ,\label{a26a}
\end{eqnarray}
where
\begin{eqnarray*}
 \varphi_{p_Dn} \left(\widetilde{\eta}, 
     x_{\perp}, z, p_z\right)&=&\varphi_{p_Dn}
    (\widetilde{\eta}, z, p_z) \varphi_n (x_{\perp})\quad , \\[0.3cm]
 \varphi_{p_Dn}
    \left(\widetilde{\eta}, z, p_z\right) &= &\frac{1}{\sqrt{2\pi}}\,
     e^{-ip_z z} \varphi_{p_Dn} (x^0, p_z)\quad .
\end{eqnarray*}
The functions $\Delta^{\mp,a,p}_Q$ have the form of the corresponding GF
in EM background, where $\tilde{\eta}$ is the time, $z$ is the
coordinate along the electric field, the mass
$m^2_Q=a^2M^2+p^2_D\frac{(bM)^2}{\rho^2}$, and the potential of the
electromagnetic field is $A_z=\tilde{\eta}\frac{\rho}{q}$. The
functions  
\begin{equation}\label{a27}
 \psi (\widetilde{\eta}, z) =
      \int^{+\infty}_{-\infty} \varphi_{p_Dn}
       \left(\widetilde{\eta}, z, p_z\right)c_{np_z}\, dp_z
\end{equation}
with  some coefficient $c_{np_z}$, obey the equation
\begin{equation}\label{a28}
\left(\left(i\partial_{\widetilde{\eta}}\right)^2 -
     \left(i\partial_z - \rho \widetilde{\eta}\right)^2 -
        {\bf P}^2_{\perp} - m_Q^2\right)
       \varphi_Q(\widetilde{\eta},z)=0\quad .
\end{equation}
Its solutions can be found \cite{NaNi} in the form of decompositions in
full sets of solutions of some first order equations in the light-cone
variables, $z_{\mp}=\tilde{\eta}\mp z$, 
\begin{eqnarray}\label{a29}
&& {}^{-}_{+}\varphi (\widetilde{\eta}, z,p_-) = \frac{1}{\sqrt{4\pi}}\,
     \rho^{-1/4}\,\exp \left\{\frac{i\rho}{2}
       \left(\frac{1}{2} z^2_{-} - z^2\right) - i \,
         \frac{p_{-}}{2} \,z_{+}\,+\nu\ln
          (\mp\widetilde{\pi}_{-})\right\},\nonumber\\[0.3cm]
&&\pi_{-} = \sqrt{\rho} \,\widetilde{\pi}_{-} = p_{-} - \rho z_ -\quad ,
     \quad \ln (\mp\widetilde{\pi}_-) = \ln (\widetilde{\pi}_{-})
      + i\pi\theta(\pm \widetilde{\pi}_{-})\quad ,\nonumber\\[0.3cm]
&&^{+}\varphi (\widetilde{\eta},z,p_{-}) = \theta(\widetilde{\pi}_{-})\,
     {_{+}\varphi} (\widetilde{\eta}, z, p_{-}) g(_+|^+)
    \quad ,\nonumber\\[0.3cm] 
&&{_{-}\varphi} (\widetilde{\eta}, z,p_-) = \theta (-\widetilde{\pi}_{-})
        ^{-}\varphi (\widetilde{\eta}, z, p_{-}) g(_-|^-)^*\quad ,
\end{eqnarray}
and $\pm$ classification has the same meaning as one in (\ref{a13}).   
Such decompositions have the
form of integrals over the parameters $p_{-}$.
Then one can present the integrals (\ref{a25}) by means of those $p_-$ integrals,
\begin{eqnarray}\label{ad1}
&&
\mp\Delta^{\pm}_Q =
     \int^{+\infty}_{-\infty}  \theta (\mp\widetilde{\pi}_{-})\,
{}^-_+\widetilde{f}dp_-\;,\\
&&{}^-_+\widetilde{f}=i\sum_n\;
{}^-_+\varphi(\widetilde{\eta},z, p_-)\varphi_n(x_{\perp})
{}^-_+\varphi^*(\widetilde{\eta}',z', p_-)
\varphi^*_n(x'_{\perp})\;.\nonumber
\end{eqnarray}
Taking into account that
$\left|g(_-|^+)\right|^2= e^{-\pi\lambda},$ one gets
\begin{eqnarray}\label{ad2}
&&\Delta^{a}_Q =-
     \int^{+\infty}_{-\infty}  \theta (-\widetilde{\pi}_{-})\,
{}_+\widetilde{f}dp_-\;,\nonumber\\
&&\Delta^{p}_Q =-
     \int^{+\infty}_{-\infty}  \theta (\widetilde{\pi}_{-})\,
{}^-\widetilde{f}dp_-\;.
\end{eqnarray}
It was shown in \cite{GGS} that the $p_-$ integrals  can always be transformed
into integrals over the proper-time $s$. After the summation over $n$,
which can be done similarly to the case $d=4$ \cite{GGS}, one gets a
result which can be written as 
\begin{eqnarray}\label{(a30)}
&& {\pm\Delta^{\mp}_Q}= 
\Delta^c_Q \pm
      \theta(\mp y_0) \Delta_Q \quad ,\\[0.3cm]
&& \Delta^c_{Q} = \int_{\Gamma_c} \,f_Q 
     ds\quad , \; 
f_Q=f_Q\left(\widetilde{\eta},x_{\perp}, \widetilde{\eta'},
             x'_{\perp}, z-z', p_D,s\right)\;, \label{a31} \\[0.3cm]
&& \Delta_Q = \epsilon (y_0) \
     \int_{\Gamma_c-\Gamma_2-\Gamma_1} \,f_Q ds
      \quad , \quad \epsilon (y_0) = \mbox{sgn} (y_0)\quad ,
\label{a32} \\[0.3cm]
&& \Delta^a_Q = \int_{\Gamma_a} \,f_Q ds + 
\theta    (z' - z) \,\int_{\Gamma_3+\Gamma_2-\Gamma_a} 
         f_Q ds \; , \label{a33} \\
&& \Delta^p_Q = \int_{\Gamma_a} \,f_Q ds + 
\theta    (z - z') \,\int_{\Gamma_3+\Gamma_2-\Gamma_a} 
         f_Q ds \; , \label{a33a} 
\end{eqnarray}
where $\theta (0)=1/2$.

\begin{figure}[h]
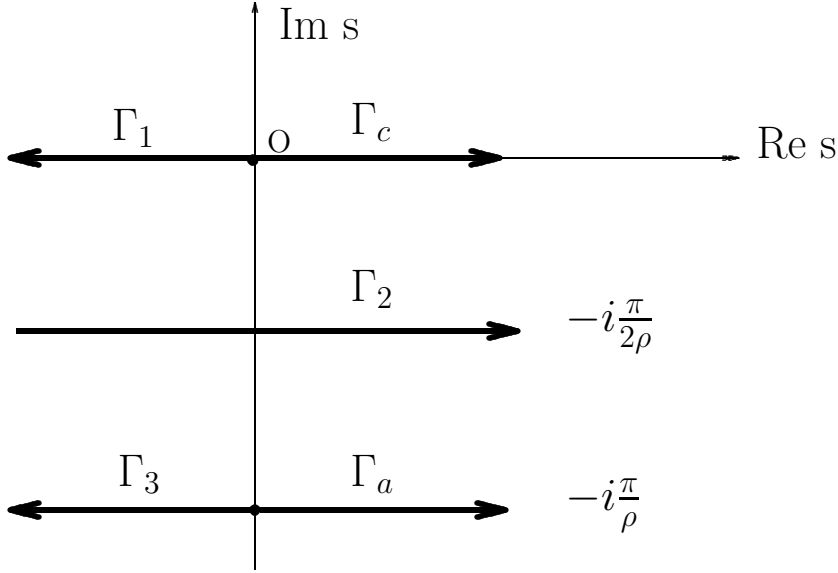

\font\thinlinefont=cmr5
\begingroup\makeatletter\ifx\SetFigFont\undefined%
\gdef\SetFigFont#1#2#3#4#5{%
  \reset@font\fontsize{#1}{#2pt}%
  \fontfamily{#3}\fontseries{#4}\fontshape{#5}%
  \selectfont}%
\fi\endgroup%
\mbox{\beginpicture
\setcoordinatesystem units <0.50000cm,0.50000cm>
\unitlength=0.50000cm
\linethickness=1pt
\setplotsymbol ({\makebox(0,0)[l]{\tencirc\symbol{'160}}})
\setshadesymbol ({\thinlinefont .})
\setlinear
%
%
\linethickness=2pt
\setplotsymbol ({\makebox(0,0)[l]{\tencirc\symbol{'161}}})
\ellipticalarc axes ratio  0.078:0.078  360 degrees 
        from 10.181 19.630 center at 10.103 19.630
%
%
\linethickness=2pt
\setplotsymbol ({\makebox(0,0)[l]{\tencirc\symbol{'161}}})
\ellipticalarc axes ratio  0.078:0.078  360 degrees 
        from 10.238 10.319 center at 10.160 10.319
%
%
\linethickness= 0.500pt
\setplotsymbol ({\thinlinefont .})
\putrule from  3.810 19.685 to 22.701 19.685
%
%
\linethickness= 0.500pt
\setplotsymbol ({\thinlinefont .})
\putrule from  3.810 19.685 to 22.701 19.685
%
%
\linethickness= 0.500pt
\setplotsymbol ({\thinlinefont .})
\putrule from 22.225 19.685 to 23.019 19.685
%
%
\plot 22.765 19.622 23.019 19.685 22.765 19.748 /
%
%
%
\linethickness= 0.500pt
\setplotsymbol ({\thinlinefont .})
\putrule from 10.160 22.860 to 10.160 23.812
%
%
\plot 10.224 23.559 10.160 23.812 10.097 23.559 /
%
%
%
\linethickness= 0.500pt
\setplotsymbol ({\thinlinefont .})
\putrule from 10.160 23.654 to 10.160  8.731
%
%
\linethickness= 0.500pt
\setplotsymbol ({\thinlinefont .})
\putrule from 22.066 19.685 to 22.860 19.685
%
%
\plot 22.606 19.622 22.860 19.685 22.606 19.748 /
%
%
%
\linethickness=2pt
\setplotsymbol ({\makebox(0,0)[l]{\tencirc\symbol{'161}}})
\putrule from  3.810 15.081 to 17.145 15.081
%
%
\plot 16.383 14.891 17.145 15.081 16.383 15.272 /
%
%
%
\linethickness=2pt
\setplotsymbol ({\makebox(0,0)[l]{\tencirc\symbol{'161}}})
\putrule from 10.160 10.319 to  3.651 10.319
%
%
\plot  4.413 10.509  3.651 10.319  4.413 10.128 /
%
%
%
\linethickness=2pt
\setplotsymbol ({\makebox(0,0)[l]{\tencirc\symbol{'161}}})
\putrule from 10.160 19.685 to 16.669 19.685
%
%
\plot 15.907 19.494 16.669 19.685 15.907 19.876 /
%
%
%
\linethickness=2pt
\setplotsymbol ({\makebox(0,0)[l]{\tencirc\symbol{'161}}})
\putrule from 10.319 10.319 to 16.828 10.319
%
%
\plot 16.066 10.128 16.828 10.319 16.066 10.509 /
%
%
%
\linethickness=2pt
\setplotsymbol ({\makebox(0,0)[l]{\tencirc\symbol{'161}}})
\putrule from 10.160 19.685 to  3.651 19.685
%
%
\plot  4.413 19.876  3.651 19.685  4.413 19.494 /
%
%
%
\put{\SetFigFont{12}{14.4}{\rmdefault}{\mddefault}{\updefault}O} [lB] at 10.478 
19.844
%
%
\put{\SetFigFont{17}{20.4}{\rmdefault}{\mddefault}{\updefault}Re s} [lB] at 23.495 
19.685
%
%
\put{\SetFigFont{17}{20.4}{\rmdefault}{\mddefault}{\updefault}Im s} [lB] at 10.795 
22.860
%
%
\put{\SetFigFont{17}{20.4}{\rmdefault}{\mddefault}{\updefault}$-i 
\frac{\pi}{\rho}$} [lB] at 18.415 10.319
%
%
\put{\SetFigFont{17}{20.4}{\rmdefault}{\mddefault}{\updefault}$-i 
\frac{\pi}{2\rho}$} [lB] at 18.415 15.081
%
%
\put{\SetFigFont{17}{20.4}{\rmdefault}{\mddefault}{\updefault}$\Gamma_1$} [lB] at  
6.350 20.161
%
%
\put{\SetFigFont{17}{20.4}{\rmdefault}{\mddefault}{\updefault}$\Gamma_c$} [lB] at 
12.700 20.320
%
%
\put{\SetFigFont{17}{20.4}{\rmdefault}{\mddefault}{\updefault}$\Gamma_2$} [lB] at 
12.700 15.875
%
%
\put{\SetFigFont{17}{20.4}{\rmdefault}{\mddefault}{\updefault}$\Gamma_3$} [lB] at  
6.509 10.954
%
%
\put{\SetFigFont{17}{20.4}{\rmdefault}{\mddefault}{\updefault}$\Gamma_a$} [lB] at 
12.700 10.954
\linethickness=0pt
\putrectangle corners at  3.581 23.838 and 23.495  8.706
\endpicture}
\caption[f1]{\label{f1}{Contours of integration 
$\Gamma_1,\Gamma_2,\Gamma_3,\Gamma_c,\Gamma_a$}}
\end{figure}

\begin{eqnarray}
&& f_Q  = \exp\{-iq\int^x_{x'}\, A^{\perp}_{\mu} dx^{\mu}\} f^{\parallel}_Q 
         \left(\widetilde{\eta}, \widetilde{\eta'}, z-z', p_D, s\right)
            f_{\perp} \left(x_{\perp}, {x'}_{\perp}, s\right) \quad , 
\label{a34}\\
&&   f_{\perp} (y_{\perp} , s) = c_d {\prod^{(d-2)/2}_{j=1}\limits} \,
        \left(\frac{qH_j}{\sin(qH_js)}\right) ,\quad \;\;\;\;\;\;\;d\mbox{ is 
even},
\nonumber\\[0.3cm]
&&   f_{\perp} (y_{\perp} , s) = c_ds^{-1/2} {\prod^{(d-3)/2}_{j=1}\limits} \,
        \left(\frac{qH_j}{\sin(qH_js)}\right),\quad  
d\mbox{ is odd}, \label{a36}   \\
&&  c_d=(4\pi)^{-d/2} \,
\exp\left\{-\frac{i}{4}\left[\pi (d-4)+ y_{\perp} qF^{(H)}  \coth (qF^{(H)}s)
       y_{\perp}\right]\right\}\; ,\nonumber \\
&& f^{\parallel}_Q \left(\widetilde{\eta}, \widetilde{\eta'}, z, p_D, s\right)
     =\frac{\rho}{\sinh(\rho s)}\,\exp\left\{-i\frac{\rho}{2}
(\widetilde{\eta} +
      \widetilde{\eta'}) z- im^2_Q s\,\right. \nonumber \\[0.3cm]
&&\left. +i\frac{\rho}{4}\left[z^2 -(\widetilde{\eta} - \widetilde{\eta'})^2 
       \right]\coth (\rho s)\right\}
\quad ,\label{a37}
\end{eqnarray}
whereas the following relations take place
\begin{eqnarray}\label{a38}
&&i\frac{d}{ds} f_Q  = \left(m^2_Q - P^2_Q\right) \,
      f_Q \quad ,\\
&&\hat{P}^2_Q = (i\partial_{\widetilde\eta})^2 - \left(i\partial_z -
      \rho \widetilde\eta\right)^2 - \hat{\bf P}^2_{\perp}\quad ,  \nonumber
\end{eqnarray}
\begin{equation}\label{a39}
\lim_{s\rightarrow +0} f_Q =
          i\delta(\widetilde{\eta} - \widetilde{\eta'}) \,
             \delta (z-z') \delta(y_{\perp})\quad.
\end{equation}

In accordance with (\ref{a24}) one can calculate the Gaussian
integrals over $p_D$ and $z$ for all the points $s$ on the contours
Fig.1. As a result one gets  
\begin{eqnarray}
&& \Delta^c (x,x') = \int_{\Gamma_c} f(x, x',s) ds\quad ,\label{a40}\\[0.3cm]
&& \Delta (x,x') = \epsilon (y_0) \int_{\Gamma} f(x,x',s) ds\quad 
,\label{a41}\\[0.3cm]
&&\Delta^a (x,x')  = - \Delta^{(1)} (x,x')
     -\Delta^{(2)}(x,x')\quad ,\label{a41a}\\[0,3cm]
&&\Delta^p (x,x')  = - \Delta^{(1)} (x,x')
     +\Delta^{(2)}(x,x')\quad ,\label{a41b}\\[0,3cm]
&& \Delta^{(1)} (x,x') = - \frac{1}{2}\,
     \int_{\Gamma_3+\Gamma_2+\Gamma_a}\, f(x,x',s) ds \quad ,\label{a42}\\    
&& \Delta^{(2)}(x,x') = \int_{\Gamma_3+\Gamma_2-\Gamma_a} f_r (x,x',s) 
    ds\quad , \label{a42a}
\end{eqnarray}
where
\begin{eqnarray}\label{a46}
&& f_r (x,x',s) =  \frac{1}{2\sqrt{\pi}}\gamma 
      \left(\frac{1}{2},\alpha\right)f(x,x',s)\quad ,\nonumber \\[0.3cm]
&& \alpha = e^{-i \pi/2} \,\frac{1}{4s(bM)^2 \omega}
    \left[\left(x_0 + {x'}_0\right) s (bM)^2 + qEy^D\right]^2\quad ,
\end{eqnarray}
and $\gamma 
      \left(\frac{1}{2},\alpha\right)$ is the incomplete 
gamma-function. Here
\begin{eqnarray}\label{a43}
&&f(x, x',s) = e^{iq\Lambda} f_{\parallel} \left(x_0,
      {x'}_0, y^D, s\right) \,f_{\perp}(y_{\perp},s)\quad ,\\
&& f_{\parallel} \left(x_0, {x'}_0, y^D, s\right) = 
    \frac{\rho}{{\rm sinh}(\rho s) \omega^{1/2}}
      \exp\left\{i\frac{qE}{2} \left(x_0 + {x'}_0\right) y^D -i
      \frac{\rho}{4} \left(x_0- {x'}_0\right)^2 {\rm coth} (\rho s) 
       \right. \nonumber\\[0.3cm]
&& \left. -i(aM)^2s +
     i \frac{\rho}{4\omega} y^2_D \,{\rm coth}(\rho s) -
        \frac{i}{4\omega} \left[(bM)^2 s \left(x_0+ {x'}_0\right)^2
          + 2qEy^D \left(x_0 +
          {x'}_0\right)\right]\right\}\quad ,\nonumber\\[0.3cm] 
&& \omega = \frac{(bM)^2}{\rho} s\,{\rm coth} (\rho s) + 
      \frac{(qE)^2}{\rho^2}\quad .\label{a45}
\end{eqnarray}
Here only $\Lambda$ depends on the choice of the gauge for the
constant field, via the integral
\begin{equation}\label{a44}
\Lambda = - \int^x_{x'} \, A_\mu dx^\mu\quad ,
\end{equation}
which is taken along the line. To get the function $f$ in an arbitrary
gauge $A'$, one has only to replace $A$ by $A'$ in the $\Lambda$.

One can see that 
\begin{eqnarray}\label{a51}
&& i \frac{d}{ds} \,f(x,x',s) = \left(M^2\Omega^2
      (x^0) - \hat{P}^2\right) f(x,x',s)\quad , \label{a52}\\[0.3cm]
&& \lim_{s\rightarrow +0} f(x,x',s) = i \delta (x-x')\quad .
\end{eqnarray} 
Thus, $f(x,x',s)$ is  Fock-Schwinger function \cite{Sch1,Fock}.  
The contour $\Gamma_c-\Gamma_2-\Gamma_1$ in (\ref{a41}) was transformed into 
    $\Gamma$ after the integration over $p_D$ and $z$. Then the
    results are consistent with the general expression for the
    commutation function obtained in \cite{GG2}.  For
    appropriate choice of gauge and in $d=4$ the function $f(x,x',s)$
    coincides with one from \cite{BO1}.

If $b\neq 0$, then the function $f(x,x',s)$ has three singular points
on the complex region $-\pi\leq\Im(\rho s)\leq 0$ which are distributed
at the imaginary axis: $\rho s_0= 0,$ $\rho s_1=-i\pi$ and $\rho
s_2=-ic_2.$ The latter point is connected with zero value of the
function $\omega$. We get an equation for $c_2$ from the condition
$\omega =0$,
\begin{equation}\label{a51.1} 
c_2\tan (c_2-\pi/2)-\left(\frac{qE}{bM}\right)^2=0\;,
\end{equation}

\begin{figure}[h]
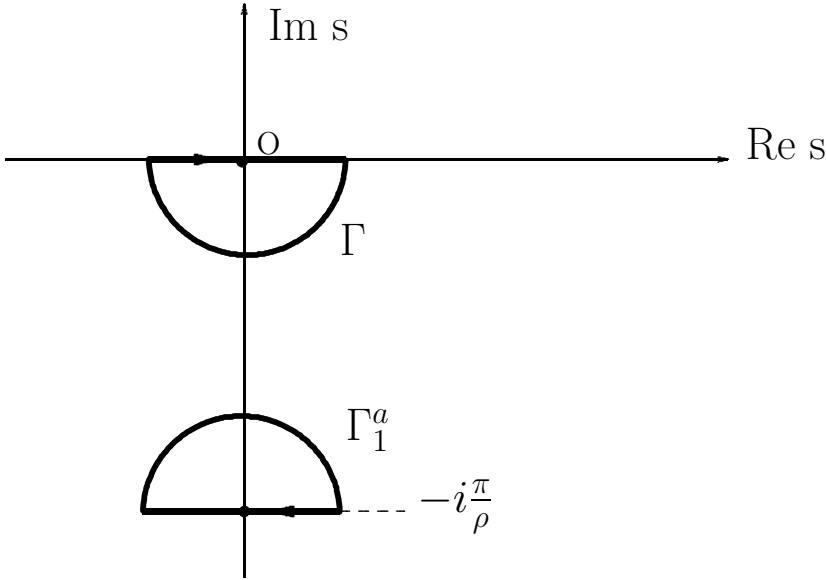

\font\thinlinefont=cmr5
\begingroup\makeatletter\ifx\SetFigFont\undefined%
\gdef\SetFigFont#1#2#3#4#5{%
  \reset@font\fontsize{#1}{#2pt}%
  \fontfamily{#3}\fontseries{#4}\fontshape{#5}%
  \selectfont}%
\fi\endgroup%
\mbox{\beginpicture
\setcoordinatesystem units <0.50000cm,0.50000cm>
\unitlength=0.50000cm
\linethickness=1pt
\setplotsymbol ({\makebox(0,0)[l]{\tencirc\symbol{'160}}})
\setshadesymbol ({\thinlinefont .})
\setlinear
%
%
\linethickness=2pt
\setplotsymbol ({\makebox(0,0)[l]{\tencirc\symbol{'161}}})
\circulararc 176.529 degrees from  7.620 19.685 center at 10.239 19.764
%
%
\linethickness=2pt
\setplotsymbol ({\makebox(0,0)[l]{\tencirc\symbol{'161}}})
\circulararc 176.575 degrees from 12.700 10.319 center at 10.082 10.240
%
%
\linethickness=2pt
\setplotsymbol ({\makebox(0,0)[l]{\tencirc\symbol{'161}}})
\ellipticalarc axes ratio  0.078:0.078  360 degrees 
        from 10.181 19.630 center at 10.103 19.630
%
%
\linethickness=2pt
\setplotsymbol ({\makebox(0,0)[l]{\tencirc\symbol{'161}}})
\ellipticalarc axes ratio  0.078:0.078  360 degrees 
        from 10.238 10.319 center at 10.160 10.319
%
%
\linethickness= 0.500pt
\setplotsymbol ({\thinlinefont .})
\putrule from  3.810 19.685 to 22.701 19.685
%
%
\linethickness= 0.500pt
\setplotsymbol ({\thinlinefont .})
\putrule from  3.810 19.685 to 22.701 19.685
%
%
\linethickness=2pt
\setplotsymbol ({\makebox(0,0)[l]{\tencirc\symbol{'161}}})
\putrule from  7.620 19.685 to 12.859 19.685
%
%
\linethickness=2pt
\setplotsymbol ({\makebox(0,0)[l]{\tencirc\symbol{'161}}})
\putrule from  7.461 10.319 to 12.700 10.319
%
%
\linethickness= 0.500pt
\setplotsymbol ({\thinlinefont .})
\putrule from 10.160 23.495 to 10.160  8.572
%
%
\linethickness= 0.500pt
\setplotsymbol ({\thinlinefont .})
\putrule from 22.225 19.685 to 23.019 19.685
%
%
\plot 22.765 19.622 23.019 19.685 22.765 19.748 /
%
%
%
\linethickness=1pt
\setplotsymbol ({\makebox(0,0)[l]{\tencirc\symbol{'160}}})
\putrule from  8.890 19.685 to  9.366 19.685
%
%
\plot  8.858 19.558  9.366 19.685  8.858 19.812 /
%
%
%
\linethickness=1pt
\setplotsymbol ({\makebox(0,0)[l]{\tencirc\symbol{'160}}})
\putrule from 11.430 10.319 to 10.954 10.319
%
%
\plot 11.462 10.446 10.954 10.319 11.462 10.192 /
%
%
%
\linethickness= 0.500pt
\setplotsymbol ({\thinlinefont .})
\putrule from 10.160 22.860 to 10.160 23.812
%
%
\plot 10.224 23.559 10.160 23.812 10.097 23.559 /
%
%
%
\linethickness= 0.500pt
\setplotsymbol ({\thinlinefont .})
\setdashes < 0.1270cm>
\plot 12.700 10.319 14.446 10.319 /
%
%
\put{\SetFigFont{12}{14.4}{\rmdefault}{\mddefault}{\updefault}O} [lB] at 10.478 
19.844
%
%
\put{\SetFigFont{17}{20.4}{\rmdefault}{\mddefault}{\updefault}$-i\frac \pi \rho $} 
[lB] at 14.764 10.319
%
%
\put{\SetFigFont{17}{20.4}{\rmdefault}{\mddefault}{\updefault}$\Gamma$} [lB] at 
12.700 17.145
%
%
\put{\SetFigFont{17}{20.4}{\rmdefault}{\mddefault}{\updefault}$\Gamma_1^a$} [lB] 
at 12.859 12.224
%
%
\put{\SetFigFont{17}{20.4}{\rmdefault}{\mddefault}{\updefault}Re s} [lB] at 23.495 
19.685
%
%
\put{\SetFigFont{17}{20.4}{\rmdefault}{\mddefault}{\updefault}Im s} [lB] at 10.795 
22.860
\linethickness=0pt
\putrectangle corners at  3.785 23.838 and 23.495  8.547
\endpicture}
\caption[f2]{\label{f2}{Contours of integration $\Gamma,\Gamma_1^a$}}
\end{figure}

where $\pi/2<c_2<\pi$. The position of this point depends on the ratio
$qE/(bM)$, e.g.  at $bM/(qE)\rightarrow 0$ one has $c_2\rightarrow
\pi$ and at $qE/(bM)\rightarrow 0$ one has $c_2\rightarrow
\pi/2.$ Notice, in the case $E=0$ it is convenient to put
$c_2=\pi/2+0$ because of the contour $\Gamma_2$ must be passed above
the singular point $s_2$ in the case as well.

If $b=0$, then $\omega=1$ and the function $f(x,x',s)$ has only two singular 
points
$s_0$ and $s_1$
on the above mentioned complex region. In this case 
the gauge invariant function
    $f_{\parallel}$ does not depend on on $x_0+x'_{0}$. In this
degenerate case it follows from (\ref{a24}), (\ref{a33}) and
(\ref{a33a}),
\begin{eqnarray}\label{a51.2}
&& \Delta^a(x,x') = \int_{\Gamma_a} \,f(x,x',s) ds + 
\theta    (-y^D) \,\int_{\Gamma_3+\Gamma_2-\Gamma_a} 
         f(x,x',s) ds \; , \\ 
&& \Delta^p(x,x') = \int_{\Gamma_a} \,f(x,x',s) ds + 
\theta    (y^D) \,\int_{\Gamma_3+\Gamma_2-\Gamma_a} 
         f(x,x',s) ds \; . 
\end{eqnarray}

Let us return to more interesting case $b\neq 0.$ Our aim is to
demonstrate that  function $\Delta^{(2)}(x,x')$ from (\ref{a42a}) can also be
presented via a proper-time integral with the kernel $f(x,x',s)$ as it
was  done for all other $\Delta$-functions.
To this end let us transform the contour $\Gamma_3 +\Gamma_2
-\Gamma_a$ in (\ref{a42a}) into two ones: $\Gamma_1^a$ (see FIG.2) and
$\Gamma_l + \Gamma_r$ (see FIG.3). 
\begin{figure}[h]
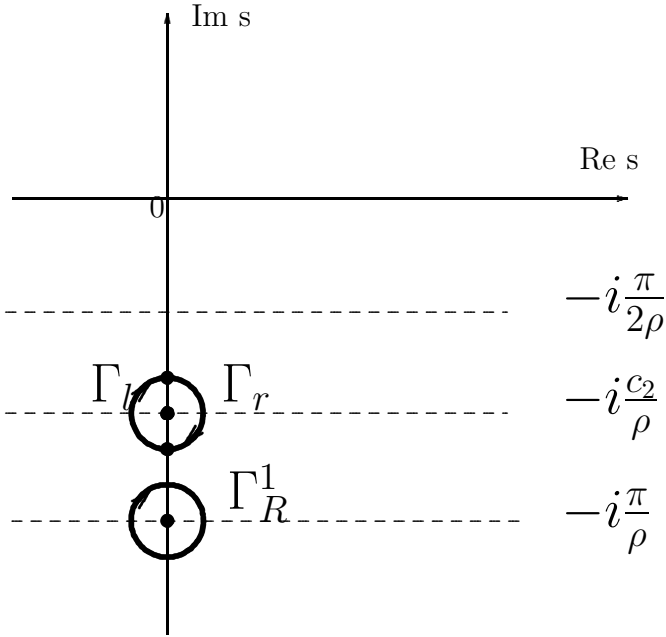

\font\thinlinefont=cmr5
\begingroup\makeatletter\ifx\SetFigFont\undefined%
\gdef\SetFigFont#1#2#3#4#5{%
  \reset@font\fontsize{#1}{#2pt}%
  \fontfamily{#3}\fontseries{#4}\fontshape{#5}%
  \selectfont}%
\fi\endgroup%
\mbox{\beginpicture
\setcoordinatesystem units <0.50000cm,0.50000cm>
\unitlength=0.50000cm
\linethickness=1pt
\setplotsymbol ({\makebox(0,0)[l]{\tencirc\symbol{'160}}})
\setshadesymbol ({\thinlinefont .})
\setlinear
%
%
\linethickness=3pt
\setplotsymbol ({\makebox(0,0)[l]{\tencirc\symbol{'162}}})
\ellipticalarc axes ratio  0.078:0.078  360 degrees 
        from  8.175 14.922 center at  8.096 14.922
%
%
\linethickness=3pt
\setplotsymbol ({\makebox(0,0)[l]{\tencirc\symbol{'162}}})
\ellipticalarc axes ratio  0.078:0.078  360 degrees 
        from  8.175 14.922 center at  8.096 14.922
%
%
\linethickness=3pt
\setplotsymbol ({\makebox(0,0)[l]{\tencirc\symbol{'162}}})
\ellipticalarc axes ratio  0.078:0.078  360 degrees 
        from  8.175 12.065 center at  8.096 12.065
%
%
\linethickness=3pt
\setplotsymbol ({\makebox(0,0)[l]{\tencirc\symbol{'162}}})
\ellipticalarc axes ratio  0.078:0.078  360 degrees 
        from  8.175 13.970 center at  8.096 13.970
%
%
\linethickness=3pt
\setplotsymbol ({\makebox(0,0)[l]{\tencirc\symbol{'162}}})
\ellipticalarc axes ratio  0.078:0.078  360 degrees 
        from  8.175 15.875 center at  8.096 15.875
%
%
\linethickness=2pt
\setplotsymbol ({\makebox(0,0)[l]{\tencirc\symbol{'161}}})
\ellipticalarc axes ratio  0.953:0.953  360 degrees 
        from  9.049 14.922 center at  8.096 14.922
%
%
\linethickness=2pt
\setplotsymbol ({\makebox(0,0)[l]{\tencirc\symbol{'161}}})
\ellipticalarc axes ratio  0.965:0.965  360 degrees 
        from  9.061 12.065 center at  8.096 12.065
%
%
\linethickness= 0.500pt
\setplotsymbol ({\thinlinefont .})
\putrule from  3.969 20.637 to 11.748 20.637
\putrule from 11.748 20.637 to 11.430 20.637
\putrule from 11.430 20.637 to 20.320 20.637
%
%
\plot 20.066 20.574 20.320 20.637 20.066 20.701 /
%
%
%
\linethickness= 0.500pt
\setplotsymbol ({\thinlinefont .})
\setdashes < 0.1270cm>
\plot  7.938  9.049  7.938  9.049 /
\plot  7.938  9.049  7.938  9.049 /
%
%
\linethickness= 0.500pt
\setplotsymbol ({\thinlinefont .})
\setsolid
\putrule from  8.096  9.049 to  8.096 25.559
%
%
\plot  8.160 25.305  8.096 25.559  8.033 25.305 /
%
%
%
\linethickness=1pt
\setplotsymbol ({\makebox(0,0)[l]{\tencirc\symbol{'160}}})
\plot  7.303 15.399  7.620 15.716 /
%
%
\plot  7.351 15.267  7.620 15.716  7.171 15.447 /
%
%
%
\linethickness=1pt
\setplotsymbol ({\makebox(0,0)[l]{\tencirc\symbol{'160}}})
\plot  7.303 12.541  7.620 12.859 /
%
%
\plot  7.351 12.410  7.620 12.859  7.171 12.589 /
%
%
%
\linethickness=1pt
\setplotsymbol ({\makebox(0,0)[l]{\tencirc\symbol{'160}}})
\plot  8.890 14.446  8.572 14.129 /
%
%
\plot  8.842 14.578  8.572 14.129  9.022 14.398 /
\linethickness= 0.500pt
\setplotsymbol ({\thinlinefont .})
\setdashes < 0.1270cm>
%
%
\plot  3.810 17.621      3.900 17.621
         3.990 17.621
         4.080 17.621
         4.169 17.621
         4.258 17.621
         4.346 17.621
         4.434 17.621
         4.521 17.621
         4.608 17.621
         4.695 17.621
         4.781 17.621
         4.867 17.621
         4.953 17.621
         5.038 17.621
         5.123 17.621
         5.207 17.621
         5.291 17.621
         5.375 17.621
         5.458 17.621
         5.541 17.621
         5.624 17.621
         5.706 17.621
         5.788 17.621
         5.869 17.621
         5.950 17.621
         6.031 17.621
         6.111 17.621
         6.191 17.621
         6.270 17.621
         6.350 17.621
         6.428 17.621
         6.507 17.621
         6.585 17.621
         6.663 17.621
         6.740 17.621
         6.817 17.621
         6.894 17.621
         6.970 17.621
         7.046 17.621
         7.121 17.621
         7.197 17.621
         7.271 17.621
         7.346 17.621
         7.420 17.621
         7.494 17.621
         7.567 17.621
         7.640 17.621
         7.713 17.621
         7.786 17.621
         7.858 17.621
         7.929 17.621
         8.001 17.621
         8.072 17.621
         8.143 17.621
         8.213 17.621
         8.283 17.621
         8.353 17.621
         8.422 17.621
         8.491 17.621
         8.560 17.621
         8.628 17.621
         8.696 17.621
         8.764 17.621
         8.831 17.621
         8.898 17.621
         8.965 17.621
         9.032 17.621
         9.098 17.621
         9.163 17.621
         9.229 17.621
         9.294 17.621
         9.359 17.621
         9.423 17.621
         9.487 17.621
         9.551 17.621
         9.615 17.621
         9.741 17.621
         9.866 17.621
         9.990 17.621
        10.112 17.621
        10.234 17.621
        10.354 17.621
        10.473 17.621
        10.590 17.621
        10.707 17.621
        10.822 17.621
        10.937 17.621
        11.050 17.621
        11.162 17.621
        11.273 17.621
        11.382 17.621
        11.491 17.621
        11.599 17.621
        11.705 17.621
        11.811 17.621
        11.915 17.621
        12.018 17.621
        12.120 17.621
        12.222 17.621
        12.322 17.621
        12.421 17.621
        12.519 17.621
        12.616 17.621
        12.713 17.621
        12.808 17.621
        12.902 17.621
        12.995 17.621
        13.088 17.621
        13.179 17.621
        13.269 17.621
        13.359 17.621
        13.448 17.621
        13.535 17.621
        13.622 17.621
        13.708 17.621
        13.793 17.621
        13.877 17.621
        13.961 17.621
        14.043 17.621
        14.125 17.621
        14.206 17.621
        14.286 17.621
        14.365 17.621
        14.444 17.621
        14.521 17.621
        14.598 17.621
        14.674 17.621
        14.750 17.621
        14.824 17.621
        14.898 17.621
        14.971 17.621
        15.044 17.621
        15.115 17.621
        15.186 17.621
        15.257 17.621
        15.326 17.621
        15.395 17.621
        15.463 17.621
        15.531 17.621
        15.598 17.621
        15.664 17.621
        15.730 17.621
        15.795 17.621
        15.860 17.621
        15.924 17.621
        15.987 17.621
        16.112 17.621
        16.235 17.621
        16.355 17.621
        16.474 17.621
        16.590 17.621
        16.705 17.621
        16.817 17.621
        16.928 17.621
        17.037 17.621
        17.145 17.621
        17.145 17.621
        /
\linethickness= 0.500pt
\setplotsymbol ({\thinlinefont .})
%
%
\plot  3.969 12.065      4.060 12.065
         4.151 12.065
         4.242 12.065
         4.332 12.065
         4.422 12.065
         4.511 12.065
         4.600 12.065
         4.688 12.065
         4.777 12.065
         4.864 12.065
         4.952 12.065
         5.039 12.065
         5.125 12.065
         5.212 12.065
         5.297 12.065
         5.383 12.065
         5.468 12.065
         5.552 12.065
         5.637 12.065
         5.721 12.065
         5.804 12.065
         5.887 12.065
         5.970 12.065
         6.052 12.065
         6.134 12.065
         6.216 12.065
         6.297 12.065
         6.378 12.065
         6.458 12.065
         6.539 12.065
         6.618 12.065
         6.698 12.065
         6.777 12.065
         6.855 12.065
         6.934 12.065
         7.012 12.065
         7.089 12.065
         7.166 12.065
         7.243 12.065
         7.320 12.065
         7.396 12.065
         7.471 12.065
         7.547 12.065
         7.622 12.065
         7.696 12.065
         7.771 12.065
         7.845 12.065
         7.918 12.065
         7.992 12.065
         8.065 12.065
         8.137 12.065
         8.209 12.065
         8.281 12.065
         8.353 12.065
         8.424 12.065
         8.495 12.065
         8.565 12.065
         8.636 12.065
         8.706 12.065
         8.775 12.065
         8.844 12.065
         8.913 12.065
         8.982 12.065
         9.050 12.065
         9.118 12.065
         9.185 12.065
         9.252 12.065
         9.319 12.065
         9.386 12.065
         9.452 12.065
         9.518 12.065
         9.584 12.065
         9.649 12.065
         9.714 12.065
         9.778 12.065
         9.843 12.065
         9.907 12.065
         9.970 12.065
        10.097 12.065
        10.222 12.065
        10.346 12.065
        10.469 12.065
        10.590 12.065
        10.711 12.065
        10.830 12.065
        10.948 12.065
        11.065 12.065
        11.180 12.065
        11.295 12.065
        11.408 12.065
        11.520 12.065
        11.631 12.065
        11.741 12.065
        11.850 12.065
        11.958 12.065
        12.065 12.065
        12.170 12.065
        12.275 12.065
        12.378 12.065
        12.481 12.065
        12.582 12.065
        12.682 12.065
        12.782 12.065
        12.880 12.065
        12.977 12.065
        13.074 12.065
        13.169 12.065
        13.263 12.065
        13.357 12.065
        13.449 12.065
        13.541 12.065
        13.631 12.065
        13.721 12.065
        13.810 12.065
        13.898 12.065
        13.985 12.065
        14.071 12.065
        14.156 12.065
        14.240 12.065
        14.324 12.065
        14.407 12.065
        14.488 12.065
        14.569 12.065
        14.650 12.065
        14.729 12.065
        14.807 12.065
        14.885 12.065
        14.962 12.065
        15.039 12.065
        15.114 12.065
        15.189 12.065
        15.263 12.065
        15.336 12.065
        15.409 12.065
        15.480 12.065
        15.552 12.065
        15.622 12.065
        15.692 12.065
        15.761 12.065
        15.829 12.065
        15.897 12.065
        15.964 12.065
        16.031 12.065
        16.097 12.065
        16.162 12.065
        16.227 12.065
        16.291 12.065
        16.417 12.065
        16.541 12.065
        16.663 12.065
        16.783 12.065
        16.901 12.065
        17.017 12.065
        17.131 12.065
        17.243 12.065
        17.354 12.065
        17.462 12.065
        17.462 12.065
        /
\linethickness= 0.500pt
\setplotsymbol ({\thinlinefont .})
%
%
\plot  3.810 14.922      3.900 14.922
         3.990 14.922
         4.080 14.922
         4.169 14.922
         4.258 14.922
         4.346 14.922
         4.434 14.922
         4.521 14.922
         4.608 14.922
         4.695 14.922
         4.781 14.922
         4.867 14.922
         4.953 14.922
         5.038 14.922
         5.123 14.922
         5.207 14.922
         5.291 14.922
         5.375 14.922
         5.458 14.922
         5.541 14.922
         5.624 14.922
         5.706 14.922
         5.788 14.922
         5.869 14.922
         5.950 14.922
         6.031 14.922
         6.111 14.922
         6.191 14.922
         6.270 14.922
         6.350 14.922
         6.428 14.922
         6.507 14.922
         6.585 14.922
         6.663 14.922
         6.740 14.922
         6.817 14.922
         6.894 14.922
         6.970 14.922
         7.046 14.922
         7.121 14.922
         7.197 14.922
         7.271 14.922
         7.346 14.922
         7.420 14.922
         7.494 14.922
         7.567 14.922
         7.640 14.922
         7.713 14.922
         7.786 14.922
         7.858 14.922
         7.929 14.922
         8.001 14.922
         8.072 14.922
         8.143 14.922
         8.213 14.922
         8.283 14.922
         8.353 14.922
         8.422 14.922
         8.491 14.922
         8.560 14.922
         8.628 14.922
         8.696 14.922
         8.764 14.922
         8.831 14.922
         8.898 14.922
         8.965 14.922
         9.032 14.922
         9.098 14.922
         9.163 14.922
         9.229 14.922
         9.294 14.922
         9.359 14.922
         9.423 14.922
         9.487 14.922
         9.551 14.922
         9.615 14.922
         9.741 14.922
         9.866 14.922
         9.990 14.922
        10.112 14.922
        10.234 14.922
        10.354 14.922
        10.473 14.922
        10.590 14.922
        10.707 14.922
        10.822 14.922
        10.937 14.922
        11.050 14.922
        11.162 14.922
        11.273 14.922
        11.382 14.922
        11.491 14.922
        11.599 14.922
        11.705 14.922
        11.811 14.922
        11.915 14.922
        12.018 14.922
        12.120 14.922
        12.222 14.922
        12.322 14.922
        12.421 14.922
        12.519 14.922
        12.616 14.922
        12.713 14.922
        12.808 14.922
        12.902 14.922
        12.995 14.922
        13.088 14.922
        13.179 14.922
        13.269 14.922
        13.359 14.922
        13.448 14.922
        13.535 14.922
        13.622 14.922
        13.708 14.922
        13.793 14.922
        13.877 14.922
        13.961 14.922
        14.043 14.922
        14.125 14.922
        14.206 14.922
        14.286 14.922
        14.365 14.922
        14.444 14.922
        14.521 14.922
        14.598 14.922
        14.674 14.922
        14.750 14.922
        14.824 14.922
        14.898 14.922
        14.971 14.922
        15.044 14.922
        15.115 14.922
        15.186 14.922
        15.257 14.922
        15.326 14.922
        15.395 14.922
        15.463 14.922
        15.531 14.922
        15.598 14.922
        15.664 14.922
        15.730 14.922
        15.795 14.922
        15.860 14.922
        15.924 14.922
        15.987 14.922
        16.112 14.922
        16.235 14.922
        16.355 14.922
        16.474 14.922
        16.590 14.922
        16.705 14.922
        16.817 14.922
        16.928 14.922
        17.037 14.922
        17.145 14.922
        17.145 14.922
        /
%
%
\put{\SetFigFont{12}{14.4}{\rmdefault}{\mddefault}{\updefault}0} [lB] at  7.620 
20.161
%
%
\put{\SetFigFont{12}{14.4}{\rmdefault}{\mddefault}{\updefault}Im s} [lB] at  8.731 
25.241
%
%
\put{\SetFigFont{12}{14.4}{\rmdefault}{\mddefault}{\updefault}Re s} [lB] at 19.050 
21.431
%
%
\put{\SetFigFont{20}{14.4}{\rmdefault}{\mddefault}{\updefault}$\Gamma_l$} [lB] at  
6.032 15.240
%
%
\put{\SetFigFont{20}{14.4}{\rmdefault}{\mddefault}{\updefault}$\Gamma_r$} [lB] at  
9.525 15.240
%
%
\put{\SetFigFont{20}{14.4}{\rmdefault}{\mddefault}{\updefault}$\Gamma^1_R$} [lB] 
at  9.684 12.383
%
%
\put{\SetFigFont{20}{14.4}{\rmdefault}{\mddefault}{\updefault}$-i\frac{\pi}{2\rho}
$} [lB] at 18.574 17.621
%
%
\put{\SetFigFont{20}{14.4}{\rmdefault}{\mddefault}{\updefault}$-i\frac{c_2}{\rho}
$} [lB] at 18.574 14.922
%
%
\put{\SetFigFont{20}{14.4}{\rmdefault}{\mddefault}{\updefault}$-i\frac{\pi}{\rho}$
} [lB] at 18.574 11.906
\linethickness=0pt
\putrectangle corners at  3.793 25.584 and 20.345  9.023
\endpicture}
\caption[f2a]{\label{f2a}{Contours of integration $\Gamma^1_R$,
$\Gamma_l$ and $\Gamma_r$}}
\end{figure}
The radius of the contour
$\Gamma_1^a$ tends to zero. The contour $\Gamma_l + \Gamma_r$ is a
infinitesimal radius clockwise circle around the singular point
$s_2.$ However, it is convenient to present it as a combination of two
semicircles $\Gamma_l$ and $ \Gamma_r$ placed on the left and the
right sides of the imaginary axis respectively.
The argument $\arg s'$ of the $\Gamma_l$ radius is in the interval 
$\pi/2\leq\arg s'\leq 3\pi/2$ and of the $\Gamma_r$ radius is in the interval 
$-\pi/2\leq\arg s'<\pi/2.$
 Then (\ref{a42a}) can
be rewritten in the form
\begin{eqnarray}
&& \Delta^{(2)}(x,x') = \int_{\Gamma_l+\Gamma_r} f_r (x,x',s) 
    ds+ r(x,x')\quad ,\label{a51.3} \\
&& r(x,x')=\int_{\Gamma_1^a} f_r (x,x',s) ds\;.\label{a51.3a}
\end{eqnarray}   
 Taking into account (\ref{Ab1}) one gets 
$$
 f_r \left(x,x',s' -i\frac{\pi}{\rho}\right) \, 
  {\raisebox{-0.6em}{$\stackrel{\textstyle\longrightarrow}
   {\scriptstyle s'\rightarrow 0}$}}\; \mbox{const}\cdot  
    \exp\{-\frac{i}{4s'} \left(x_0 - {x'}_0\right)^2 \} \quad .
$$
Hence, one can see that $r(x,x')=0, \;\partial_0 r(x,x')=0$ at any
$x_0-x_0'$. 
Moreover, using (\ref{a51}), it is easy to see that the
    distribution $r(x,x')$ obeys the equation (\ref{a1}). Thus,
    $r(x,x')$ is equal to zero identically. 
The function $f_r(x,x's)$ is $2\pi$ periodic function of the argument
    $\arg s'$ of the  $\Gamma_l$  and $\Gamma_r$ radiuses. One needs
to    take into account
 the asymptotic decomposition (\ref{Aa5}) which is valid in the region
$-3\pi/2<\arg \alpha <3\pi/2$. Then, using (\ref{Aa5}) one gets from (\ref{a51.3})
\begin{equation}\label{a51.4}
\Delta^{(2)}(x,x')=\frac{1}{2}\left\{
\begin{array}{ll}
-\int_{\Gamma_l+\Gamma_r}f(x,x's)ds\;, &-5\pi/4<\beta <-3\pi/4,\\
-\int_{\Gamma_l-\Gamma_r}f(x,x's)ds\;, &-3\pi/4\leq\beta <-\pi/4,\\
\int_{\Gamma_l+\Gamma_r}f(x,x's)ds\;, &-\pi/4\leq\beta \leq\pi/4,\\
\int_{\Gamma_l-\Gamma_r}f(x,x's)ds\;, &\pi/4<\beta \leq 3\pi/4,
\end{array}
\right.
\end{equation}
where $\beta=\arg \left[(x_0+x'_0)c_2(bM)^2(-i)+\rho qEy^D\right].$

One can verify that  expression (\ref{a51.4}) is continuous in the
boundaries of the $\beta$ intervals. Then, using  (\ref{a52}), one can
demonstrate that the  representation (\ref{a51.4}) obeys the equation (\ref{a1}).
One can also verify that the representation $\Delta^{(1)}(x,x')$ (\ref{a42})
obeys the equation (\ref{a1}). Thus, all the $\Delta$-Green functions considered
here, excluding those marked by the index ``c'', are solutions of the
equation (\ref{a1}). The important difference between basic Green functions
$\Delta^c(x,x')$,  $\Delta^{(1)}(x,x')$ and
$\Delta(x,x')$,  $\Delta^{(2)}(x,x')$
is that  the first ones are symmetric under simultaneous  change of sign in
$x_0,\;x'_0,\;x_D,\;x'_D$ and the seconds ones  change sign in this case.

Note finally that using proper-time kernel $f(x,x',s)$ (\ref{a43}) one
can easily construct Schwinger out-in effective action
\[
\Gamma_{out-in}=-i\int dx\int_0^{\infty}s^{-1}\,f(x,x,s)\,ds;.
\]
Similar out-in effective action (but in another approximation) will be
discussed in Section IV for scalar electrodynamics. As regards to
in-in effective action its representation is more complicated
\cite{FG,FGS} and will be discussed in the next publication.

\section{Vacuum instability and back reaction of  particles created}

All the information about the  processes   of particles creation, annihilation, 
and
scattering in an external field (without radiative corrections) 
can be  extracted from the matrices   
$G\left({}_{\zeta}|{}^{\zeta'}\right)$  (\ref{e10}). These matrices  define a  
canonical transformation between in and out creation and annihilation
operators in the generalized Furry representation \cite{Git1,FGS},
\begin{eqnarray}\label{e10b}
&&a^{\dagger}(out)=a^{\dagger}(in)G\left({}_{+}|{}^{+}\right)+
b(in)G\left({}_{-}|{}^{+}\right),\nonumber\\
&&-b(out)=a^{\dagger}(in)G\left({}_{+}|{}^{-}\right)+
b(in)G\left({}_{-}|{}^{-}\right).
\end{eqnarray}
 Here 
$a_{l}^{\dagger}(in)$, $b_{l}^{\dagger}(in)$, $a_{l}(in)$, $b_{l}(in)$
 are  creation and annihilation operators of in-particles and
antiparticles respectively  
 and $a_{l}^{\dagger}(out)$,$b_{l}^{\dagger}(out)$,
 $a_{l}(out),\;b_{l}(out)$ are ones of out-particles and
antiparticles, $l$ are possible quantum numbers (in our case $l=p_D,n$) . For 
example, 
the mean numbers  of  particles  created
(which are also equal to the numbers of pairs created) 
by the external field from the in-vacuum $|0,in>$  with a given 
quantum number $l$ is
\begin{equation}\label{e11}
 N_{l}= <0,in|a_{l}^{\dagger}(out)a_{l}(out)|0,in>=
\left|g\left({}_{-}|{}^{+}\right)
\right|^2.
\end{equation} 
(for a review of gravitational particles creation, see
\cite{GMM,Par}.) 
The standard space coordinate
 volume regularization was used to get the latter formula, so that $\delta ( p_j-
p'_j)\rightarrow \delta_{p_j,p'_j} $.
The probability  for a vacuum to remain a vacuum is
\begin{equation}\label{e13}
P_v=|c_v|^2  =\exp\left\{- \sum_{l}\ln\left(
1+ N_{l}\right) \right\}\;,
\end{equation}
where $|0,out>$ is the out-vacuum.

Remember that we are discussing  the case in which the electric field
acts for an infinite time. However, one can analyse the problem in
finite times $T=x^0_{out}-x^0_{in}$, acting similar to  
 \cite{GG1}. Then the mean numbers of $(p_D,n)$- particles created by the
external field are
\begin{equation}\label{a14}
N_{p_D n} = |g(-|^+)|^2 = e^{-\pi\lambda},\;
\;   \mbox{if}\quad \sqrt{\rho}\,T\gg  1,\;\;\mbox{and}\;\sqrt{\rho} T\gg
   \lambda, \; \mbox{and} \; \rho^2 T\gg |qEp_D|\;, 
\end{equation}
where $\lambda$ is defined in (\ref{a12}).
The latter conditions take  place  for large  $T$. 

At d=4  (\ref{a14}) coincides with the one obtained in
\cite{SD}, and  at $b=0$ (the gravitational field is absent) it
coincides with the one obtained in  \cite{GG1}.
 $(bM)^2/\rho^2<1,$ and the number $N_{p_Dn}$ depends
 from  $p^2_D$ more weaker than from other quantum numbers
${\bf P}^2_{\perp}$. 

If the condition $p_D^2(bM)^2/\rho^3<<1$ takes place (the
gravitational field is in a sense weaker than the electric one) the
$p_D$ dependence of the mean numbers (\ref{a14}) is similar to the
pure electrodynamical  case. Thus  \cite{GG1} one can
evaluate that $\int dp_D=(eE)^{-1}\rho^2T$. Then one
can estimate the particle creation per unit of time  similarly to 
\cite{SD}. In strong enough gravitational fields time dependence of
the effect is nonlinear one and  demands a special study. 
  
To get the total number $N$ of  particles created one has to sum  over 
the quantum numbers $n, p_D$. The  sum over the momenta can be 
 easily transformed into an integral. Thus, if $b=0$ one gets
result presented in \cite{GG1}. If $b\neq 0$ and $d=4$ the total
number of pairs created per space coordinate volume has the form
\begin{equation}\label{add1}
\tilde{n}=\frac{\sum_{p_D,n}N_{p_Dn}}{\int d{\bf x}}=
\frac{
\beta(1)}{8\pi^{2}}
\frac{\rho^{3/2}}{bM}\exp\left\{-\pi\frac{(aM)^2}{\rho}\right\},
\end{equation}
where 
$$
\beta(n)=
\frac{qH}{\sinh (n\pi qH/\rho)}\;.$$
The observable number density of the created pairs in the asymptotic
region $x_0=x_0^{out}\rightarrow\infty$ is given by
\begin{equation}\label{add1a}
{n}^{cr}=\tilde{n}/\Omega^3(x_0)\;.
\end{equation}
These  results coincide with ones in \cite{SD}.

In case $b\rightarrow 0$ the expression (\ref{add1}) is growing
unlimited. In this case the particles are created in main by the
electric field, whereas the parameter $b$ plays a role of ``cut-off''
factor, which eliminates creation of particles with extremely high 
momenta along the electric field. One can see that from the expression
(\ref{a14}). Thus, the limit $b\rightarrow 0$ corresponds to the case
of the electric field which acts an infinite time when the number of
particles created is proportional to the time of the field action. In
fact, we are interesting in the case when the field action time $T$ is
big enough to obey the stabilization condition, which has the form
$T>>(qE)^{-1/2}$ for intense field. As was already remarked above, in
this case $\int dp_D=(qE)^{-1}\rho^2T.$ Then it is clear that 
parameter $b$ has to be interpreted as a quantity which is inversely
proportional to the field action time $T$ making the substitution
$b^{-1}\rightarrow TM(\pi\sqrt{\rho})^{-1}.$

The vacuum-to-vacuum transition probability 
   can be calculated, using   formula (\ref{e13}). 
Thus, we get an analog of the well-known Schwinger  formula \cite{Sch1}
in the case under consideration. For the case $b=0$ the result was
   given in \cite{GG1}. For the case $b\neq 0$ and $d=4$ one gets
\begin{equation}\label{add3}
P_v=\exp\left\{-\mu \tilde{n}\int d{\bf x}
\right\},\;\; 
\mu=\sum_{l=0}^{\infty}\frac{(-1)^{l}\beta (l+1)}
{(l+1)^{3/2}\beta (1)}
\exp \left\{-l\pi
\frac{(aM)^2}{\rho}\right\}\;.
\end{equation}
This result  coincides with the one obtained in \cite{BO1}.

Let operator of  current of scalar field operator $\phi(x)$ obeying the
Klein-Gordon equation (\ref{a1}) has  a form
\begin{equation}\label{3.76}
j_{\mu}=q\left.\left(\hat{P}_\mu+\hat{P'}_\mu^*\right)
\frac{1}{2}\left[\phi^{\dagger}(x'),\;\phi (x)\right]_+\right|_{x=x'}\;,
\end{equation}
where $\,\hat{P}_\mu = i\frac{\partial}{\partial x^\mu} - q\,A_\mu (x)$
and  $\,\hat{P'}_\mu^* = -i\frac{\partial}{\partial x'^\mu} - q\,A_\mu
(x')$, and  operator of Chernikov-Tagirov 
 metric energy-momentum tensor (EMT) of the field operator \cite{CT}  reads
\begin{eqnarray}
&&T_{\mu\nu}=T^{can}_{\mu\nu}+t_{\mu\nu}\;,\label{3.77}\\
&& T^{can}_{\mu\nu}=\left.B_{\mu\nu}
\frac{1}{2}\left[\phi^{\dagger}(x'),\;\phi (x)\right]_+\right|_{x=x'}
\;,\label{3.78}\\
&&B_{\mu\nu}=
\hat{P'}_\mu^*\hat{P}_\nu+\hat{P'}_\nu^*\hat{P}_\mu-
\eta_{\mu\nu}\left(\hat{P'}_\mu^*\hat{P}^\mu-
M^2\Omega^2(x_0)\right)\;,\label{3.79}\\
&&t_{\mu\nu}=C_{\mu\nu}
\frac{1}{2}\left[\phi^{\dagger}(x),\;\phi (x)\right]_+
\;,\label{3.80}\\
&&C_{\mu\nu}=-\frac{1}{3}\left(\partial_{\mu}\partial_{\nu}-\eta_{\mu\nu}
\partial_{\lambda}\partial^{\lambda}\right)\;,\label{3.81}
\end{eqnarray}
where $T_{\mu\nu}^{can}$ is canonical EMT operator. Here $\phi (x)$ is scalar,
$j_\mu$ is vector, and $T_{\mu\nu}$ is tensor under proper homogeneous Lorentz
transformation.
For $d=4$ one can get  operators of scalar field  $\Phi(x)$, 
the current vector $J_\mu$ and EMT  $\tau_{\mu\nu}$ under
general coordinate transformation by using formulas
\begin{equation}\label{3.76a}
\Phi(x)=\Omega^{-1}(x^0)\phi(x),\;\;
J_{\mu}=\Omega^{-2}(x^0)j_{\mu},\;\;\tau_{\mu\nu}=\Omega^{-2}(x^0)T_{\mu\nu}.
\end{equation} 
 We are going to
discuss the following matrix elements with these operators 
\begin{eqnarray}
&&<j_{\mu}>^c=<0,out|j_{\mu}|0,in>c_v^{-1}\;,\label{3.82}\\
&&<T_{\mu\nu}>^c=<0,out|T_{\mu\nu}|0,in>c_v^{-1}\;,\label{3.83}\\
&&<j_{\mu}>^{in}=<0,in|j_{\mu}|0,in>\;,\label{3.84}\\
&&<T_{\mu\nu}>^{in}=<0,in|T_{\mu\nu}|0,in>\;,\label{3.85}\\
&&<j_{\mu}>^{out}=<0,out|j_{\mu}|0,out>\;,\label{3.86}\\
&&<T_{\mu\nu}>^{out}=<0,out|T_{\mu\nu}|0,out>\;.\label{3.87}
\end{eqnarray}
Using GF which were  found before,  one can present  these matrix
  elements in the following form,
\begin{eqnarray}
&&<j_{\mu}>^c=q\left.\left(\hat{P}_\mu+\hat{P'}_\mu^*\right)
(-i)\Delta^c(x,x')\right|_{x=x'}\;,\label{3.88}\\
&& <T_{\mu\nu}>^c=\left.B_{\mu\nu}(-i)\Delta^c(x,x')
\right|_{x=x'}+C_{\mu\nu}(-i)\Delta^c(x,x)
\;,\label{3.89}\\
&&<j_{\mu}>^{in}=<j_{\mu}>^{c}+<j_{\mu}>^{(1)}+<j_{\mu}>^{(2)}\;,\;\;
\;\label{3.90a}\\
&&<j_{\mu}>^{out}=<j_{\mu}>^{c}+<j_{\mu}>^{(1)}-<j_{\mu}>^{(2)}\;,\label{3.90b}\\
&&<T_{\mu\nu}>^{in}=<T_{\mu\nu}>^{c}+<T_{\mu\nu}>^{(1)}+<T_{\mu\nu}>^{(2)}
\;,\;\;\;\label{3.91a}\\
&&<T_{\mu\nu}>^{out}=<T_{\mu\nu}>^{c}+<T_{\mu\nu}>^{(1)}-<T_{\mu\nu}>^{(2)}
\;,\label{3.91b}\\
&&<j_{\mu}>^{(1,2)}=q\left.\left(\hat{P}_\mu+\hat{P'}_\mu^*\right)
(-i)\Delta^{(1,2)}(x,x')\right|_{x=x'}\;,\label{3.92}\\
&& <T_{\mu\nu}>^{(1,2)}=\left.B_{\mu\nu}(-i)\Delta^{(1,2)}(x,x')
\right|_{x=x'}+C_{\mu\nu}(-i)\Delta^{(1,2)}(x,x)
\;,\label{3.93}
\end{eqnarray}
where  GF are given by  eq. (\ref{a40}), (\ref{a42}) and
(\ref{a42a}), and the relation
$$\Delta^c(x,x)=\frac{1}{2}\left[\Delta^-(x,x)-\Delta^+(x,x)\right]$$
is used.

The components $<j_{\mu}>^{(1,2)}$    and  $<T_{\mu\nu}>^{(1,2)}$
  in the expressions (\ref{3.90a}) and (\ref{3.91a}) can not be
calculated in the frame of the perturbation theory with respect to the
external background or in the frame of WKB method. Among them only the term
$<j_{\mu}>^{in}$ was calculated before and only in the pure electric field in 
flat space ($b=0$), see \cite{FGS}. The only expression (\ref{3.89}) for
$<T_{\mu\nu}>^c$ has to be regularized and renormalized. The
  expression (\ref{3.88}) for term
$<j_{\mu}>^c$ is finite after the regularization lifting. The terms
$<j_{\mu}>^{(1,2)}$ and $<T_{\mu\nu}>^{(1,2)}$ are also finite. That
is consistent with the fact that the ultraviolet divergences have a local
nature and result (as in the theory without external field) from the
leading local terms at $s\rightarrow +0$. The nonzero contributions to
the expressions $<j_{\mu}>^{(1,2)}$ and $<T_{\mu\nu}>^{(1,2)}$ are
related to global features of the theory and indicate the vacuum instability.

Let us introduce the normalized values of current and EMT (which maybe
easily connected with observable values), 
\begin{eqnarray}
&&j_{\mu}^{cr}=\tilde{j}_{\mu}^{cr}/\Omega^3 (x_0)\;,\label{3.95}\\
&&T_{\mu\nu}^{cr}=\tilde{T}_{\mu\nu}^{cr}/\Omega^3 (x_0)\;,\label{3.96}
\end{eqnarray}
(in some convenient asymptotic 
region $x_0=x_0^{as}$), where the 
corresponding densities per space-coordinates volume are
\begin{eqnarray}
&&\tilde{j}_{\mu}^{cr}=\frac{\int d{\bf x}\left(<j_{\mu}>^{in}
-<j_{\mu}>^{out}\right)}{\int d{\bf x}}\;,\;\;x_0=x_0^{as},\label{3.97}\\
&&\tilde{T}_{\mu\nu}^{cr}=\frac{\int d{\bf x}\left(<T_{\mu\nu}>^{in}
-<T_{\mu\nu}>^{out}\right)}{\int d{\bf x}}\;,\;\;x_0=x_0^{as}\,,\label{3.98}
\end{eqnarray}
according to the definitions (\ref{3.84}) - (\ref{3.87}).

Then, using representations (\ref{3.90a}) - (\ref{3.93}) one gets from
(\ref{3.97}) and (\ref{3.98}),
\begin{eqnarray}
&&\tilde{j}_{\mu}^{cr}=2<j_{\mu}>^{(2)}\;,\;\;x_0=x_0^{as},\label{3.99}\\
&&\tilde{T}_{\mu\nu}^{cr}=2<T_{\mu\nu}>^{(2)}\;,\;\;x_0=x_0^{as}.\label{3.100}
\end{eqnarray}

To study the backreaction of particles created on the electromagnetic 
field and metrics one needs the expressions $<j_{\mu}>^{in}$ and 
$<T_{\mu\nu}>^{in}$  at all times $x_0$. Below we are going to analyse these
expressions for different times. Note, that the functions $\Delta^c(x,x')$
(\ref{a40}) and $\Delta^{(1)}(x,x')$ (\ref{a42}) at $x=x'$
 are even functions on $x_0$. Thus, the functions $<T_{\mu\nu}>^c$
 and $<T_{\mu\nu}>^{(1)}$
are also even ones and do not vanish at $x_0\rightarrow 0$, whereas
the functions $<j_{\mu}>^c$ and $<j_{\mu}>^{(1)}$ are odd ones and vanish in this
limit. Moreover, we have $<j_\mu>^c=0$ for all $x_0$ at $b=0$. 

The proper-time integral $\Delta^{(2)}(x,x')$ (\ref{a42a})
 is a odd function on $x_0$ at $x=x'$
and vanishes at $x_0\rightarrow 0$. Thus, the expression
$<T_{\mu\nu}>^{(2)}$
 is also a
odd function on $x_0$ and vanishes in this limit. The term $<j_{\mu}>^{(2)}$ is an
even function on $x_0$ and is different from zero in this limit if
$E\neq 0$.

One can see, using the expressions obtained for  GF,  that
all off-diagonal matrix elements of EMT are equal to zero:
\begin{equation}\label{3.102a}
<T_{\mu\nu}>^{c}=<T_{\mu\nu}>^{(1)}=<T_{\mu\nu}>^{(2)}=0,\;\mu\neq\nu;
\end{equation}
and only $x^D$ components (along the electric field) of the currents are
different from zero and vanish in the absence of the electric field.

Below we are going to analyse contributions to the physical quantities
(\ref{3.99}) and (\ref{3.100}) for $d=4$ at $x_0=x_0^{as}$, which are stipulated
by the GF 
$\Delta^{(2)}(x,x')$.

Using the asymptotic form of $\Delta^{(2)}(x,x')$ given in
(\ref{Aa7}), one can see that at $x_0>>\sqrt{\rho}/(bM)$
 the following  asymptotic expression takes place
\begin{equation}\label{3.101}
<j_{\mu}>^{(2)}=q^2|E|\rho^{-1}\tilde{n}^{cr}\delta_\mu^3.
\end{equation}
Taking into account eq. (\ref{Aa4a}) one gets an asymptotic
expression for diagonal $<T_{\mu\nu}>^{(2)}:$ 
\begin{eqnarray}\label{3.102}
&&<T_{00}>^{(2)}=\left[\rho^2x_0^2+a^2M^2+qH\coth\left(\pi
qH/\rho\right)\right]\frac{\tilde{n}^{cr}}{\rho x_0}\;,\nonumber\\
&&<T_{11}>^{(2)}=<T_{22}>^{(2)}=
qH\coth\left(\pi
qH/\rho\right)\frac{\tilde{n}^{cr}}{2\rho x_0}\;,\nonumber\\
&&<T_{33}>^{(2)}=\left[(qEx_0)^2+\frac{\rho^3}{2\pi (bM)^2}
\right]\frac{\tilde{n}^{cr}}{\rho x_0}\;,
\end{eqnarray}
where $\tilde{n}^{cr}$ is given by (\ref{add1}).

Doubling the expression (\ref{3.101}) and (\ref{3.102}) 
according the eq. (\ref{3.99}) and (\ref{3.100}), one
gets the mean densities for current and EMT of particles created. It
turns out that these quantities are proportional to the density of
total number of particles and antiparticles  created  $2\tilde{n}^{cr}$ for the
infinite time and do not change their structure  with increasing of $x_0$. The
latter means that one can consider all the expressions obtained at any
fixed $x_0$ if $x_0\gg \sqrt{\rho}/bM$. For a strong background
$a^2M^2/\rho\leq 1$ and
therefore this time has to obey the condition $x_0>>a/b$. Thus, in the strong
background our asymptotic conformal time $x_0$ corresponds to the
large cosmological time $t$.

Note, that one can neglect the second term in the brackets of the
expression (\ref{3.102}) for $<T_{33}>^{(2)}$
 at $bM/(qE)\leq 1$. Also one can neglect both the term $a^2M^2$
in the brackets of the
expression (\ref{3.102}) for
 $<T_{00}>^{(2)}$ in case of strong external background
$a^2M^2/\rho\leq 1$ 
and  third term
in the same expression if the magnetic field strength is not big
enough $qH/\rho\leq 1.$  The current density $
\tilde{j}_{\mu}^{cr}=2<j_{\mu}>^{(2)}$ does
not depend of the asymptotic time. At $b\rightarrow 0$ this expression coincides
with one for  flat space $
<\tilde{j}_{\mu}>^{cr}=2|q|\tilde{n}^{cr}\delta_\mu^3.$ The pressure component 
along the
electric field direction  
$\tilde{T}_{33}^{cr}=2<T_{33}>^{(2)}$
is growing with time upon the action of the
field. However, if $qE/(bM)<<1$ then the asymptotic condition for  $x_0$ is
consistent with the fact that the term $(qEx_0)^2$ in the expressions 
$<T_{33}>^{(2)}$ from (\ref{3.102})
will not  be dominant until big enough time  $x_0$. In this
case one can neglect the contribution which depends on the electric
field if the field is switched off before. Note, that more accurate
analysis of back-reaction of created particles to current requires the
numerical estimations (compare with purely EM case, \cite{KES}).

The components of the pressure in the directions perpendicular to the
electric field $\tilde{T}_{11}^{cr}=2<T_{11}>^{(2)}
$ and $\tilde{T}_{22}^{cr}=2<T_{22}>^{(2)}
$ are growing in relation to the  $\tilde{T}_{33}^{cr}$ when the
magnetic field is increasing. However, a very strong magnetic field
decreases all the components of $\tilde{T}_{\mu\nu}^{cr}$
 and $\tilde{j}_{\mu}^{cr}$ because  the particles
creation will be decreased.  

One can remark that the asymptotic behaviour of $<j_{\mu}>^{(1)}$ and
$<T_{\mu\nu}>^{(1)}$ is defined by the asymptotic expression
for $\Delta^{(1)}(x,x')$ (\ref{Aa10}). Then one gets
\begin{equation}\label{3.103}
<j_{\mu}>^{(1)}=<j_{\mu}>^{(2)}\,,\;\;\;<T_{\mu\nu}>^{(1)}=<T_{\mu\nu}>^{(2)}\;.
\end{equation}

The expression (\ref{3.88}) does not need to be renormalized, thus,
one can easily verify (using (\ref{Aa1}) and (\ref{Aa2})) that the
relation 
$<j_{\mu}>^{c}\sim x_0^{-1}\rightarrow 0$ holds asymptotically.
An estimation of the finite part of  
$<T_{\mu\nu}>^{c}$ can be made only after  renormalization. We are
going to consider this problem, together with others, related to the
renormalization, in the next paper. 

To get an idea about the behaviour of the expressions (\ref{3.90a})
and  (\ref{3.91a}) at finite time  let us estimate their
components for some small $x_0$, namely for $x_0^2<<\rho/(bM)^2.$
Dislocating the contour $\Gamma_2$ to the real axis and neglecting the
divergent terms in $<T_{\mu\nu}>^{c}+<T_{\mu\nu}>^{(1)}$ which do not
depend on the background, one can see that 
\begin{eqnarray}
&&<j_{\mu}>^{in}=\Re 
<j_{\mu}>^{c}+<j_{\mu}>^{(2)}+<j_{\mu}>^{(3)}\;,\label{3.104}\\
&&<T_{\mu\nu}>^{in}=\Re
<T_{\mu\nu}>^{ren}+<T_{\mu\nu}>^{(2)}+<T_{\mu\nu}>^{(3)}\;.\label{3.105}
\end{eqnarray}
Here $
<T_{\mu\nu}>^{ren}$ is obtained from $
<T_{\mu\nu}>^{c}$ as a result of such a procedure and
\begin{eqnarray}
&&<j_{\mu}>^{(3)}=q\left.\left(\hat{P}_\mu+\hat{P'}_\mu^*\right)
(-i)\Delta^{(3)}(x,x')\right|_{x=x'}\;,\label{3.106}\\
&& <T_{\mu\nu}>^{(3)}=\left.B_{\mu\nu}(-i)\Delta^{(3)}(x,x')
\right|_{x=x'}+C_{\mu\nu}(-i)\Delta^{(3)}(x,x)
\;,\label{3.107}
\end{eqnarray}
where $\Delta^{(3)}(x,x')$ is defined in (\ref{Aa8}). As was already
remarked, $<j_{\mu}>^{c}$ and $<j_{\mu}>^{(3)}$ are odd functions on
$x_0$. They  vanish at $x_0\rightarrow 0.$ That is why the leading term in
(\ref{3.104}) is defined by $<j_{\mu}>^{(2)}.$ Using the expressions
(\ref{Ab4}), obtained in the Appendix B, one gets
\begin{equation}\label{3.108}
<j_{\mu}>^{(2)}=q^2n^{(2)}\delta_\mu^3.
\end{equation}
 
If $bM/(qE)\geq 1$, then the odd function $ <T_{\mu\nu}>^{(2)}$
vanishes at  $x_0\rightarrow 0$ and the leading contribution to
(\ref{3.105}) is defined by $\Re
<T_{\mu\nu}>^{ren}$ and $<T_{\mu\nu}>^{(3)}.$
However, if $bM/(qE)<<1,$ in the domain where
$qE/(bM)^2>>x_0^2>>(qE)^{-1},$ the situation is different and the
leading contributions can come from $ <T_{\mu\nu}>^{(2)}.$ That
happens since the expressions $\Re
<T_{\mu\nu}>^{ren}$ and $<T_{\mu\nu}>^{(3)}$ remain finite at
$b\rightarrow 0.$ Using (\ref{Ab4}) and (\ref{Ab5}) one can find
expression for diagonal $<T_{\mu\nu}>^{(2)},$
\begin{eqnarray}\label{3.109}
&&<T_{00}>^{(2)}=\sum_{l=1,2,3}<T_{ll}>^{(2)}+a^2M^2x_0\pi 
\left(\frac{bM}{qE}\right)^2\tilde{n}^{cr}\;,\nonumber\\
&&<T_{11}>^{(2)}=<T_{22}>^{(2)}=qH\coth (\pi H/E)
x_0\pi 
\left(\frac{bM}{qE}\right)^2\tilde{n}^{cr}
\;,\nonumber\\
&&<T_{33}>^{(2)}=5qEx_0\tilde{n}^{cr}\;.
\end{eqnarray}
>From it one can see that $ <T_{33}>^{(2)}$ and $ <T_{00}>^{(2)}$ are
growing unlimited at $b\rightarrow 0$ because of increasing of 
$\tilde{n}^{cr}.$

The expression (\ref{3.108}) at $bM/(qE)<<1$ coincides with the
asymptotic one (\ref{3.101}). That demonstrates that in the quasi-flat
metrics, when the particles are created mainly due to the
electric field only, any time  far enough ($x_0+T/2>>
(qE)^{-1/2}$) from the initial time  $(x_0^{in}=-T/2),$ can be
considered as the asymptotic time. However, that is not true for the
EMT $ <T_{\mu\nu}>^{(2)}$ and the expression  (\ref{3.109}) differs
essentially from the asymptotic one  (\ref{3.102}). That happens since
the vacuum definition for the particles with big longitudinal momenta
$(p_3\geq qEx_0),$ differs essentially from $|0,out>$ vacuum. Namely, 
those particles  mainly contribute to $ <T_{\mu\nu}>^{(2)}.$
Technically that means that one can not obtain the normal form by
means of the $\Delta^{out}$-function and, consequently, to calculate
the mean value of EMT of particles created at $x_0<<\sqrt{qE}/(bM)$ it
is not enough to know components of $ <T_{\mu\nu}>^{in}.$ Note
finally that the expressions for mean values of energy momentum tensor
may be applied for the estimations of back-reaction of created
particles to the gravitational background like it has been done in
\cite{HH}. However, such study involves the renormalization of EMT. 
It also cannot be done analytically. We will discuss
these questions in an another place.

\section{Effective action in scalar QED in FRW Universe with constant
 electromagnetic field}  

In this section (which is somehow outside of above discussion) 
we will consider $d=4$ scalar QED in curved spacetime 
with EM field. That is more complicated, interacting theory. Using
such a theory as an example we will briefly discuss how one can
generalize the results of above discussion to interacting theories. 
We again consider gravitational-electromagnetic background of weakly
curved constant curvature spacetime.
The Lagrangian of the theory may be written as
following:
\begin{eqnarray}\label{c1}
&&L=L_m+L_{\rm ext}\;, \\
&&L_m=\frac{1}{2}\left(\partial_{\mu}\phi_1-qA_{\mu}\phi_2\right)^2+\frac{1}{2}\left(
\partial_{\mu}\phi_2+qA_{\mu}\phi_1\right)^2-\frac{1}{2}M^2\phi^2+\frac{1}{2}\xi
R\phi^2-\frac{1}{4!}\lambda\phi^4-\frac{1}{4}F_{\mu\nu}F^{\mu\nu}\;,
\nonumber \\
&&L_{\rm ext}=\Lambda+\kappa
R+a_1R^2+a_2C_{\mu\nu\alpha\beta}^2+a_3G\;,
\nonumber
\end{eqnarray}
where $\phi^2=\phi_i \phi_i=\phi_1^2+\phi_2^2 $, the parameterization
of fields in terms of two real  scalars is taken, $R$ is curvature,
$C_{\mu\nu\alpha\beta}$ is Weyl tensor, $G$ is Gauss-Bonnet
combination. The necessity of introduction of $L_{\rm ext}$ is
dictated by the multiplicative renormalizability of the theory (see
\cite{BOS}). Note also that we consider the theory in curved spacetime
with EM field, hence $A_{\mu}\rightarrow A_{\mu}+\tilde{A}_{\mu} $,
where $\tilde{A}_{\mu}$ is background EM field. For simplicity, we
limit below only in constant curvature space. The well known example
of such spaces is De Sitter space which is often used for description
of the inflationary Universe (exponentially expanding one).

There are few ways to study such a theory on the quantum level. When
one treats the external background exactly, like it has been done in
previous section, the first step is to calculate scalar Green
functions.

Due to the fact that the Maxwell theory is conformally invariant, 
Green functions for EM field will be the same as in the flat
space. Then, using the proper-time representation for the Green
functions, quantum corrections to $\Gamma_{\rm out-in}$ or
$\Gamma_{\rm in-in}$ can be calculated (where $\Gamma_{\rm out-in}$,
$\Gamma_{\rm in-in}$ are effective actions for the probability
amplitudes and for the mean values respectively). Then, of course, the
external background is kept to be exact and perturbation theory over
only coupling constants $\lambda, e$ is used. However, such
calculation is extremely hard what can be already understood from very
complicated form of the Green functions in the previous
section. Indeed, these Green functions should be used instead of
standard momentum space propagators in Feynman diagrams (for an
explicit example in pure constant EM background, see
\cite{FGS}). Hence, such calculation using Green functions of the
previous section will be given in another place.

Instead, we will adopt quasi-local expansion for the effective action
here. We will make the explicit use of RG \cite{BOS} in such
calculation. Note that in quasi-local approach to effective action
(i.e. when we do not treat an external background exactly but make the
derivative expansion of the effective action over derivatives from
metric and EM field) the difference between $\Gamma_{\rm out-in}$, and
$\Gamma_{\rm in-in}$ is not seen. That is why we denote the effective
action here as $\Gamma$ simply. (Of course, particle creation
phenomenon is also hard to see in quasi-local approximation.) As
background we will consider De Sitter space with background EM field
$A_{\mu}$ (no background scalar field!).

Using the fact that the theory with the Lagrangian (\ref{c2}) is
multiplicatively renormalizable, one can write RG equation for the
effective action on the above background as following:
\begin{equation}\label{c2}
\left(\mu\frac{\partial}{\partial \mu}+\beta_i
 \frac{\partial}{\partial \lambda_i}
-\gamma_A \tilde{A}_{\mu}\frac{\partial}{\partial \tilde{A}_{\mu} } \right)
\Gamma (\mu,\lambda_i,\tilde{A}_{\mu},g_{\mu\nu})=0\;,
\end{equation}
where $\tilde{A}_{\mu}$ is background EM field chosen so that
$\tilde{F}_{\mu\nu}$ is (almost) constant (the well-known example is
constant electric field),
$\lambda_i=(q,\lambda,M,\xi,\Lambda,\kappa,a_1,a_2,a_3),\; \gamma_{A}$
is $\gamma$-function for $\tilde{A}_{\mu}$, and $\beta_i$ is
$\beta$-function for $\lambda_i$.

The solution of RG equation (\ref{c2}) can be easily found using the
method of characteristics (see \cite{CW} for flat space and \cite{EO1}
for curved spacetime). In such a way, one obtains RG improved
effective action (which makes summation of all leading logarithms of
the perturbation theory). Dropping the details which are very similar
to the ones given in ref. \cite{EO1}, we get RG improved effective
action up to the terms of second order on the curvature invariants:
\begin{equation}\label{c3}
\Gamma_{RG}=\Lambda (t)+\kappa
(t)+a_1(t)R^2+a_3(t)G-\frac{1}{4}\frac{q^2}{q^2(t)}\tilde{F}_{\mu\nu}^2\;,
\end{equation}
where the effective coupling constants are given as following (see
\cite{CW} for the flat space and \cite{EO1} for curved spacetime):
\begin{eqnarray}\label{c4}
&& q^2(t) = q^2 \left( 1- \frac{2q^2t}{3(4\pi)^2} \right)^{-1},
\
\ \ \Phi^2 (t) = \Phi^2 \left( 1- \frac{2q^2t}{3(4\pi)^2}
\right)^{-9}, \nonumber \\
&& \lambda (t) = \frac{1}{10} q^2(t) \left[ \sqrt{719} \, \tan
\left(  \frac{1}{2} \sqrt{719} \, \log q^2(t) +C \right) +19
\right], \nonumber \\
&& C = \arctan  \left[ \frac{1}{\sqrt{719} } \left( \frac{10
\lambda}{q^2} -19 \right) \right] - \frac{1}{2} \sqrt{719} \,
\log
q^2, \nonumber \\
&& M^2(t)= M^2 \left[ \frac{q^2(t)}{q^2} \right]^{-26/5}
\frac{\cos^{2/5} \left( \frac{1}{2} \sqrt{719} \, \log q^2 +C
\right) }{\cos^{2/5} \left( \frac{1}{2} \sqrt{719} \, \log q^2(t)
+C \right) }, \nonumber \\
&& \xi (t) = \frac{1}{6} + \left(\xi - \frac{1}{6} \right)
 \left[ \frac{q^2(t)}{q^2} \right]^{-26/5} \frac{\cos^{2/5}
\left(
\frac{1}{2} \sqrt{719} \, \log q^2 +C \right) }{\cos^{2/5} \left(
\frac{1}{2} \sqrt{719} \, \log q^2(t) +C \right) }, \nonumber \\
&& a_2(t) = a_2 + \frac{7 t}{60(4\pi)^2}, \ \ \ a_3(t) =a_3-
\frac{8t}{45(4\pi)^2}, \nonumber \\
&& \lambda (t) = \Lambda + M^4 A_1 (t), \ \ \ \kappa (t) = \kappa
+ 2M^2 (\xi -1/6) A_1(t), \ \ \ a_1(t) = a_1 + (\xi -1/6)^2
A_1(t),
\nonumber \\
&& A_1(t) = \int_0^t \frac{dt}{(4\pi)^2} \,  \left[
\frac{q^2(t)}{q^2} \right]^{-52/5} \frac{\cos^{4/5} \left(
\frac{1}{2} \sqrt{719} \, \log q^2 +C \right) }{\cos^{4/5} \left(
\frac{1}{2} \sqrt{719} \, \log q^2(t) +C \right) }\;.
\end{eqnarray}
Note that one should use the fact that
$C_{\mu\nu\alpha\beta}=0,\;G=\frac{R^2}{6}$ on de Sitter
background. The classical Lagrangian is used as boundary condition in
the derivation of $\Gamma_{RG}$.

Now, the question is about the choice of RG parameter $t$. We are
dealing with the theory with few effective masses. RG improvement in such
a theory is not easy due to the fact that one has generally speaking
to generalize the total mass matrix (there are two masses from the
scalar sector plus four more from EM sector, see \cite{EO1,BKMN}).

In order to find RG parameters $t$ one has to specify situation in
more details. For example, let us consider the case when the
background EM field is chosen to be constant magnetic field, i.e. 
$\frac{1}{4}\tilde{F}_{\mu\nu}\tilde{F}^{\mu\nu}=\frac{1}{2}H^2$.
Then, one can consider different limiting cases.

First, let $H\gg R$. Then all effective masses are becoming
proportional and there is unique choice for RG parameter $t:
t=\frac{1}{2}\ln \frac{qH}{\mu ^2}$. In this case, the curvature
effects are not relevant, $\Gamma_{RG}$ is given by its last term in
(\ref{c4}). That is the case actually discussed by Schwinger
\cite{Sch1}.

Second, let $R\gg H$. Then, the unique choice for $t$ is 
$t=\frac{1}{2}\ln \frac{R/4}{\mu ^2}$. Now, EM field effects are not
relevant. We get purely gravitational effective action discussed in
ref. \cite{BOS,BO2,EO1}.

When both fields are of the same order the choice for $t$ (in order to
make summation of leading logarithms) is $t=\frac{1}{2}\ln
\frac{R/4+qH}{\mu ^2}$. If, in addition, we choose the initial values
for mass and $\xi$ as following: $M^2=0,\;\xi=\frac{1}{6}$, then we
get 
\begin{equation}\label{c5}
\Gamma_{RG}=-\frac{8t}{270(4\pi)^2}R^2-\left(1-\frac{2q^2t}{3(4\pi)^2}\right)\frac
{H^2}{2}\;,
\end{equation}
where only H-dependent terms kept. The expression (\ref{c5})
generalizes the well-known effective Lagrangian for magnetic field
obtained by Schwinger \cite{Sch1} to curved spacetime.  Note that in (\ref{c5}) 
the
curvature effects have been taken into account.

It is known that in flat space \cite{Sch1} the Schwinger effective
Lagrangian leads to the stationary point $H\neq 0$ (due to quantum
corrections). However, this stationary point is known to be the
maximum of the effective action. Hence, it does not lead to the
possibility of new ground state with account of quantum
corrections. In curved space, one can analyze $\Gamma_{RG}$ (\ref{c5})
using equation of motion $\frac{\partial \Gamma _{RG}}{\partial
H}$. The result of this analysis shows that position of the flat-space
maximum $H\neq 0$ is moving due to curvature corrections. However, the
nature of this stationary point is the same (it is the maximum of the
effective action). Similarly, one can analyze (now numerically) the
case with $M^2\neq 0$ and an arbitrary $\xi$. However, we expect that
existence of new ground state may occur only in the regime with strong curvature 
(where an
external gravitational field is treated exactly like in the previous section).

\section{Conclusion}
In summary, we considered massive scalar field in expanding radiation
dominated FRW Universe filled by the constant EM field. Taking
scalar-gravitational coupling constant to be equal to its conformal
value and making conformal transformation, the theory is formulated as
flat space theory with time-dependent mass and in external EM
field. For such a theory, which is generally speaking theory with
unstable vacuum, we found proper-time representation for all Green
functions (i.e. for in-in, out-in, and so on Green functions). Note
that proper-time representation (over a finite contour in proper-time
complex plane) for in-in Green function in the combination of the
external gravitational and EM fields is given here for the first time.

As some applications, we discuss particles creation by the external
fields in arbitrary dimensions. Combined action of gravitational and
EM fields may produce new interesting features in this phenomenon, in
particular, affect the rate of particles creation.

Using the proper-time representation for Green functions, the
proper-time representation for the effective actions is also found
(or, in other words, vacuum polarization is found) keeping external
background exactly. 

Let us discuss one more possible application of the Green functions
obtained-the calculation of radiative corrections to the transition
amplitudes and to mean values of different quantities in the
interacting theory. The corresponding general theory (Furry picture
for theories with unstable vacuum) was developed in
\cite{Git1,FG,FGS,BFG}. As an example, let us consider scalar
self-interacting theory with the interaction
$-\frac{1}{4!}(\phi^+\phi)^2$. The same background as in Sect. II will
be considered. Let us imagine that one has to calculate first-order
radiative corrections to the two-point Green function which serves for
calculation of mean values. Schematically,
we have to consider the following diagram;

\begin{figure}[h]
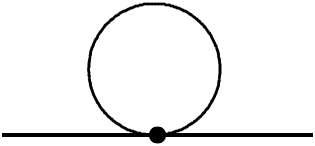

\font\thinlinefont=cmr5
\begingroup\makeatletter\ifx\SetFigFont\undefined%
\gdef\SetFigFont#1#2#3#4#5{%
  \reset@font\fontsize{#1}{#2pt}%
  \fontfamily{#3}\fontseries{#4}\fontshape{#5}%
  \selectfont}%
\fi\endgroup%
\mbox{\beginpicture
\setcoordinatesystem units <0.50000cm,0.50000cm>
\unitlength=0.50000cm
\linethickness=1pt
\setplotsymbol ({\makebox(0,0)[l]{\tencirc\symbol{'160}}})
\setshadesymbol ({\thinlinefont .})
\setlinear
%
%
\linethickness=4pt
\setplotsymbol ({\makebox(0,0)[l]{\tencirc\symbol{'163}}})
\ellipticalarc axes ratio  0.080:0.080  360 degrees 
        from 10.558 17.780 center at 10.478 17.780
%
%
\linethickness=1pt
\setplotsymbol ({\makebox(0,0)[l]{\tencirc\symbol{'160}}})
\ellipticalarc axes ratio  1.748:1.748  360 degrees 
        from 12.145 19.526 center at 10.397 19.526
%
%
\linethickness=1pt
\setplotsymbol ({\makebox(0,0)[l]{\tencirc\symbol{'160}}})
\putrule from  6.350 17.780 to 14.605 17.780
\linethickness=0pt
\putrectangle corners at  6.303 21.304 and 14.652 17.621
\endpicture}

\caption[f3]{\label{f3}{A radiative correction to the propagator}}
\end{figure}
with $\Delta_{in}$ functions. Hence, these corrections are
proportional to in-in propagator in
coinciding points, i.e. $\sim \Delta _{in}^c(x,x)$. Or more
completely, above-drawn diagram is proportional to the expression
\[
\sim \int \Delta (y,x')\Delta (x',x')\Delta (x',z)dx'\;.
\]
The divergent part of the diagram is well-known, it is the same as in
flat space and is proportional (in dimensional regularization) to 
$\frac{\lambda M^2}{(n-4)}$. The finite part of the diagram can be
easily found, using explicit form of the proper-time representation
for the corresponding Green function, given in Sect. II. Of course, 
this finite part is different from one in the absence of the external
background. To calculate the finite radiative corrections to the causal
propagator, one has to take the same diagram with $\Delta^c$ 
functions (or with out-in Green functions). The divergent part of the
diagram will be the same, however, the finite part is different from
the previous case (as it follows from the explicit structure of Green
functions, presented in Sect. II). In the same way one can analyse
other radiative corrections.

Using explicit form of Green functions in Sect. II, we calculate the
effective actions (vacuum polarization) in proper-time representation.
The knowledge of out-in effective action gives an alternative way to
define the particles creation a la' Schwinger \cite{Sch1} (for an
explicit example see \cite{BOS}). The in-in effective action can be used
to study the back reaction of the particles created to the external
background. Such an analysis is not easy and will be presented in
another place where also a generalization of results of Sect. II for
an arbitrary $\xi$ will be done. For example, using explicit form of
in-in GF in proper-time representation it could be of interest also to
construct proper-time representation for in-in effective action.

In Sect. IV we, presented another approach to the effective action
(derivative expansion of effective action) in the external
gravitational-EM background. Scalar QED is considered as an example;
RG improved effective action (up to the terms of second order on
curvature and EM strength) is calculated on constant curvature weakly
curved spacetime with weak constant EM field. Such an effective action
gives the extension of the well-known Schwinger effective Lagrangian,
taking into account curvature effects. It may be also applied to the
study of back reaction of quantum field theory to external background.

Finally, similar technique may be applied to analyze the behaviour of
spinor fields in gravitational-EM background. The calculation of all
the Green functions in such a theory, using proper-time
representation, may be the necessary step in the study of chiral
symmetry breaking in QED and in the four-fermion models under the
action of gravitational and EM fields. Such a study may have an
immediate important application to early Universe, for example,
through the construction of inflationary Universe where role of
inflaton is played by the condensate $<\bar{\Psi}\Psi>$. One can also
analyse symmetry breaking phenomenon under the combined  action of
gravitational and EM fields in the Standard Model (using also its
gauged NJL form \cite{NJL}), or Grand Unified
Theories in the same way as it has been done in curved spacetime
(without EM field) \cite{BOS}.  

Note also that GF investigation developed in this paper maybe
extremly useful for the study of Casimir effect due to combined
action of 
gravitational and electromagnetic fields (for an introduction to
Casimir effect in pure EM or pure gravitational case, see \cite{30}).

\section{Acknowledgments}

D. Gitman, S. Gavrilov and S. Odintsov are grateful to Brazilian foundations   
CNPq  
and  FAPESP respectively for support. Besides, S. Gavrilov thanks Brazilian
foundation CAPES for support, and S. Odintsov thanks Russian 
Foundation of Fundamental Research for partial support  under the Grant No 
96-02-16017.

\appendix

\section{ Asymptotics of  $\Delta$- functions}

Let us calculate the asymptotic behavior of 
$\Delta^{(1)}(x,x')$ and $\Delta^{(2)}(x,x')$ in the case
$x^2_0>>\rho/(bM)^2,$ and $x\rightarrow x'$ at $d=4.$

First, it is useful to take into account the following formulas
\begin{eqnarray}
&&\hat{P}_\mu e^{iq\Lambda} = e^{iq\Lambda}i\partial_{\mu}\;,\;\;
\hat{P'}_\mu^* e^{iq\Lambda}=
e^{iq\Lambda}(-i\partial'_{\mu})\;,\label{Aa1}\\
&&\hat{P}_{j}f(x,x',s)=  \hat{P'}_{j}^*f(x,x',s)\;,\;\; j=1,2,3\;,\label{Aa2}\\
&&\left.\hat{P}_0 \hat{P'}_0^* f(x,x's)\right|_{x=x'}=\left.\left[
\hat{P}_0^2+2\left(\omega^{-1}(bM)^2sx_0\right)^2-i\omega^{-1}(bM)^2s
\right]
f(x,x',s)\right|_{x=x'}\,.
\label{Aa3}
\end{eqnarray} 
Then, using eq.(\ref{Aa3}) and equation (\ref{a52}) one gets
\begin{eqnarray}\label{Aa4}
&&\left.\hat{P}_0 \hat{P'}_0^* f(x,x',s)\right|_{x=x'}=\nonumber\\
&&\left.\left[
-i\frac{d}{ds}+
\hat{P}_{j}^2+M^2\Omega^2(x_0)+2\left(\omega^{-1}(bM)^2sx_0\right)^2-\omega^{-1}i(
bM)^2s
\right]
f(x,x',s)\right|_{x=x'}\,.
\end{eqnarray}
If the $\Delta$-function obeys the equation (\ref{a1}) then the
action of the operator 
$-i\frac{d}{ds}$ on this function is equal to zero.
Further we are going to use the formula (\ref{Aa4})
 for simplification of the calculation.
 Besides, one can see that
for  such kind of functions the following relation holds,
\begin{equation}\label{Aa4a}
\left. \left(\hat{P'}_\mu^*\hat{P}^\mu -M^2\Omega^2(x_0) \right)
 \Delta^{(\ldots)}(x,x')\right|_{x=x'}=2\partial^2_0
\Delta^{(\ldots)}(x,x)\,.
\end{equation}
So, if one uses these formulas in calculations it is enough to find GF
asymptotics for the case $x_0-x'_0=0$.

Let us find the asymptotic behavior of $\Delta^{(2)}(x,x')$ function
given by the eq. (\ref{a42a}).
If $b\neq 0$ and  $x_0-x'_0=0$ the kernel $f_r(x,x's)$ has no
singular point $ s_1=-i\pi/\rho$. 
Below the line of contours $\Gamma_3-\Gamma_a$ 
next singular point of this function is $ s_2 - i\pi/\rho.$ Thus, one
can shift the line of the composite contour $\Gamma_3-\Gamma_a$ below
along the imaginary axis until the neighborhood of the point
$ s_2 - i\pi/\rho.$ The contour which is obtained in such a way from
 $\Gamma_3+\Gamma_2-\Gamma_a$ can be closed on the parts $\Re
s\rightarrow\pm\infty$ and then can be transformed into a closed
contour which includes the points $s_1$ and $s_2$ as well. At the same time
this contour can be situated far enough from the point $s_1$, so that
$\left|\rho s \omega^{-1}\right|$ is always not zero at it. Then one
can use the asymptotic decomposition
\cite{HTF}
\begin{equation}\label{Aa5}
\gamma\left(1/2,\alpha\right) =\sqrt{\pi}-  e^{-\alpha}
\alpha^{-1/2}\left[1 + O (\alpha^{-1})\right]\;,\;\;x_0>0\;.
\end{equation}
The contribution from the first term of eq. (\ref{Aa5}) at the contour
in question is exponentially small since $\Re (-i\rho s
\omega^{-1})<0.$ The rest terms of eq. (\ref{Aa5}) form a series in
inverse powers of $x_0^2 (bM)^2/\rho$, the corresponding functions in
the decomposition coefficients 
$f(x,x',s)e^{-\alpha}
     \alpha^{-1/2}\left[1 + O (\alpha^{-1})\right]
$
 have only one singular point $s_1$ (the pole) inside the contour. The
contributions from these terms  can be estimated tightening the contour
to the point $s_1.$ The first one of these terms defines the leading
contribution into the asymptotics of $\Delta^{(2)}(x,x'),$ 
\begin{equation}\label{Aa6}
\Delta^{(2)}(x,x') =\frac{1}{2\sqrt{\pi}}\int_{\Gamma^1_R}f(x,x',s)  e^{-\alpha}
     \alpha^{-1/2}ds
          \quad , \\[0.3cm]
\end{equation}
where the contour $\Gamma_R^1$ (see FIG.3) is  
a circle with infinitesimal radius around the singular point $s_1.$
Calculating the residue, one gets an expression for
$\Delta^{(2)}(x,x')$ which defines the leading asymptotics in
$<j_{\mu}>^{(2)}$ and $<T_{\mu\nu}>^{(2)},$
\begin{equation}\label{Aa7}
\Delta^{(2)}(x,x') =i\frac{\tilde{n}^{cr}}{\rho (x_0+x'_0)}
\exp\{iq\Lambda + \frac{i}{2}qE(x_0+x'_0)y^3-\frac{\rho^3(y_3)^2}{4\pi
(bM)^2}+\frac{1}{4}y_{\perp}
qF\cot\left(\pi
qF/\rho\right)y_{\perp}
         \} \quad , \\[0.3cm]
\end{equation}
where $\tilde{n}^{cr}$ is defined in (\ref{add1}). This expression is also
valid at $x_0<0,$ since the function $\Delta^{(2)}(x,x')$ is an odd
one in $x_0$ at $x=x'.$ 

Consider  the asymptotic behavior of $\Delta^{(1)}(x,x')$ function
given by eq. (\ref{a42}). Since $\rho s \omega^{-1}$ is not equal to zero
and $\Re (-i\rho s \omega^{-1})<0$
at the contour $\Gamma_2$ the corresponding asymptotic contribution in integral 
(\ref{a42}) is exponentially small. Then one needs to evaluate only
the part of this integral given by form
\begin{equation}\label{Aa8}
\Delta^{(3)} (x,x') = - \frac{1}{2}\,
     \int_{\Gamma_3+\Gamma_a}\, f(x,x',s) ds \quad ,
\end{equation}
Let us introduce the variable $\tau ,$ $\rho s=-i\pi+\tau.$
Since  $\Re (-i\rho s \omega^{-1})<0$ at the contours $\Gamma_3$ and
$\Gamma_a,$ and since $|\omega|^{-1}$  increases monotonous with
$|\tau|,$ then the leading asymptotic contribution is defined by the
behaviour of the function $f(x,x',s)$ at small $\tau$ and has the form
\begin{equation}\label{Aa9}
\Delta^{(3)} (x,x') = e^{-i\pi/4}\pi^{-1/2}f(x,x',s_1) 
     \int_{0}^{\infty}d\tau \tau^{-1}e^{-i\tau \rho x_0^2} \quad .
\end{equation}
Calculating the integral one gets
\begin{equation}\label{Aa10}
\Delta^{(1)} (x,x') = \mbox{sign} (x_0)\Delta^{(2)} (x,x') \,
\quad ,
\end{equation}
where $ \Delta^{(2)} (x,x')$ is given eq.(\ref{Aa7}).

\section{Small time expansion of $\Delta^{(2)}$-function} 

Let calculate a small time expansion of $\Delta^{(2)}(x,x')$
function
given by eq. (\ref{a42a})
 in a case
$x^2_0<<\rho /(bM)^2$ and $x\rightarrow x'$ at $d=4.$
The small $\alpha$ expansion of the incomplete $\gamma$-function is  
valid at the contours $\Gamma_a,$ $\Gamma_2$ and $\Gamma_3$ and it 
 has a form \cite{HTF}
\begin{equation}\label{Ab1}
\gamma\left(1/2,\alpha\right) = e^{-\alpha}
     \alpha^{1/2}\left[2 +(4/3)\alpha+ 0 \left(\alpha^{2}\right)\right]
          \quad , 
\end{equation}
where second term in the square brackets is necessary for the
calculations of $\hat{P'}_0^*\hat{P}_0$ and $\hat{P'}_3^*\hat{P}_3$
actions.
Since term (\ref{a51.3a}) is equal to zero  it is convenient
to calculate the function $\Delta^{(2)}(x,x'),$ closing the contour
$\Gamma_3+\Gamma_2-\Gamma_a$ on the area 
$\Re
s\rightarrow\pm\infty$ and then tightening it to the singular point
$s_2.$ Then the leading contributions to 
$<j_{\mu}>^{(2)}$ and $<T_{\mu\nu}>^{(2)},$ are defined by the integral
\begin{equation}\label{Ab2}
\Delta^{(2)}(x,x') =\frac{1}{2\sqrt{\pi}}\int_{\Gamma_l+\Gamma_r}f(x,x',s)  
e^{-\alpha}
     \alpha^{1/2}\left[2 +(4/3)\alpha\right]ds
          \quad , \\[0.3cm]
\end{equation}
where the contour $\Gamma_l+\Gamma_r$ (see FIG.3) is  a infinitesimal radius
clockwise circle around the singular point $s_2.$
According to eq. (\ref{a51.1}) $\omega=0$ at $s=s_2.$ Then in the
neighborhood of this  singular point if $s=s_2+s'$ one gets expansion
\begin{eqnarray}\label{Ab3}
&&\omega=\omega'\rho s'+(1/2)\omega''(\rho s')^2\;,\\
&&\omega'=-i\frac{1}{c_2}\left(\frac{bM}{\rho}\right)^2\left[
c_2^2+\left(\frac{qE}{bM}\right)^2+\left(\frac{qE}{bM}\right)^4\right]\;,
\nonumber\\
&&\omega''=2\frac{1}{c^2_2}\left(\frac{bM}{\rho}\right)^2
\left[1+\left(\frac{qE}{bM}\right)^2\right]
\left[c_2^2+\left(\frac{qE}{bM}\right)^4\right]\;.\nonumber
\end{eqnarray}
Calculating residue at the point $s_2$ one finds
the small time expansion of $\Delta^{(2)}$-function,
\begin{eqnarray}\label{Ab4}
&&\Delta^{(2)}(x,x')=e^{iq\Lambda}\left\{\left[i(x_0+x'_0)c_2(bM)^2/\rho
-qEy^3\right]n^{(2)}/(qE)\varphi_0\right. \nonumber\\
&&+\left.i\left[-(x_0+x'_0)^3\frac{(bM)^4}{6qE\rho^2}K(2)+
(1/2)(x_0+x'_0)(y_3)^2qEK(0)
\right]
\right\}\,\\
&&\varphi_0=\exp
\left(i\frac{qE}{2}(x_0+x'_0)y^3-\frac{\rho}{4c_2}
\left(\frac{qE}{bM}\right)^2(x_0-x'_0)^2
-\frac{\rho^3(y_3)^2}{4c_2(bM)^2}
\right)\;,\nonumber\\
&&K(l)=-\frac{(-i)^lc_2^l}{c_2^2+\left(\frac{qE}{bM}\right)^2+
\left(\frac{qE}{bM}\right)^4}
\left\{a^2M^2/\rho-c_2^{-1}(l-1/2)-c_2^{-1}\left(\frac{qE}{bM}\right)^2
\right.\nonumber\\
&&\left.+(qH/\rho)\coth (c_2qH/\rho)+2c_2^{-1}\frac{
\left[1+\left(\frac{qE}{bM}\right)^2\right]\left[c_2^2+\left(\frac{qE}{bM}\right)^
4\right]}
{c_2^2+\left(\frac{qE}{bM}\right)^2+\left(\frac{qE}{bM}\right)^4}
\right\}\frac{\rho}{2qE}n^{(2)}\;,\nonumber\\
&&n^{(2)}=\frac{\sqrt{c_2^2+\left(\frac{qE}{bM}\right)^4}}
{8\pi^{3/2}c_2\left[
c_2^2+\left(\frac{qE}{bM}\right)^2+\left(\frac{qE}{bM}\right)^4
\right]}
\frac{q^3HE^2\rho^{3/2}}{(bM)^3\sin (c_2qH/\rho)}e^{-c_2a^2M^2/\rho}\;,\nonumber
\end{eqnarray}
If $bM/qE<<1$, coefficients $K(l)$ and $n^{(2)}$ from (\ref{Ab4}) have
more simple form, and in the case of an intensive electric field
($a^2M^2/(qE)<1,\; |H/E|<1$) one has
\begin{eqnarray}\label{Ab5}
&&n^{(2)}=\tilde{n}^{cr}\;,\nonumber\\
&&K(l)=-(-i)^{l}(1/2)\pi^l\tilde{n}^{cr}\;.
\end{eqnarray}


\newpage

\begin{thebibliography}{99}
\bibitem{Sch1} J. Schwinger, Phys. Rev. {\bf 82}, 664 (1951)  
\bibitem{BdW}B. S. DeWitt, {\em Dynamical Theory of Groups and Fields}
(Gordon and Breach, New York 1965);  Phys. Rept.C {\bf 19}, 295 (1975)
\bibitem{Git1} D.M. Gitman, J. Phys. A {\bf 10}, 2007 (1977);
S.P. Gavrilov and D.M. Gitman, Sov. Phys. Journ. No.6, 491 (1980)
\bibitem{FG} E.S. Fradkin and D.M. Gitman, Preprint MIT (1978)
pp. 1-58; Preprint KFKI-1979-83, pp. 1-105; Preprint PhIAN (P.N. Lebedev
Institute) 106 (1979) pp. 1-62; 107 (1979) 1-40;  Fortschr. Phys. {\bf 29}
(1981) 381-411
\bibitem{FGS}E.S. Fradkin, D.M. Gitman and S.M. Shvartsman, {\em
Quantum Electrodynamics with Unstable Vacuum} (Springer-Verlag, Berlin 1991)
\bibitem{BFG}I.L. Buchbinder and D.M. Gitman, Izw.VUZov Fizika
(Sov.Phys.Journ.) No 3 (1979) 90; ibid No 4 (1979) 55; ibid N7 (1979) 16; 
I.L. Buchbinder, E.S. Fradkin and D.M. Gitman, Fortschr. Phys. {\bf 29}
(1981) 187   
\bibitem{GGS} S.P. Gavrilov, D.M. Gitman and Sh.M. Shvartsman, Sov. J. Nucl. Phys.
(USA) {\bf 29}, 567 (1979); 715 (1979); Kratk.Soob.Fiz. (P.N. Lebedev
Inst.) No 2, 22 (1979); D.M. Gitman, M.D. Noskov and Sh.M. Shvartsman,
Int. J. Mod. Phys. A{\bf 6} (1991) 4437
\bibitem{TW}M.S. Turner and L.M. Widrow, Phys. Rev. {\bf D37} (1988)
2743
\bibitem{GGV}M. Gasperini, M. Giovannini and G. Veneziano,
Phys. Rev. Lett. {\bf 75} (1995) 3796  
\bibitem{SD}G. Sch\"afer and H. Dehnen, J. Phys. {\bf A13} (1980) 517
\bibitem{BO1}I.L. Buchbinder and S.D. Odintsov, Izv. VUZov Fizika
(Sov. Phys. Journ.) No.5 (1982) 12.  
\bibitem{BOS} I.L. Buchbinder, S.D. Odintsov and I.L. Shapiro,
{\em Effective Action in Quantum Gravity} (IOP Publishing, Bristol and
Philadelphia 1992)
\bibitem{BO2}I.L. Buchbinder and S.D. Odintsov, Izv. VUZov Fizika
(Sov. Phys. Journ.) N8 (1983) 50; ibid N12 (1983) 108;
Yad. Fiz. (Sov. J. Nucl. Phys.) {\bf 40} (1984) 1338; Lett. Nuovo Cim.
{\bf 42} (1985) 379; T. Muta and S.D. Odintsov,
Mod. Phys. Lett. {\bf A6} (1991) 3641
\bibitem{BKO}I.L. Buchbinder, E.N. Kirillova and S.D. Odintsov, Class.
Quant. Grav. {\bf 4} (1987) 711  
\bibitem{CW}S. Coleman and E. Weinberg, Phys. Rev. {\bf D7} (1973)
1888; M.B. Einhorn and D.R. Jones, Nucl. Phys. {\bf B211} (1983) 29;
G.B. West, Phys. Rev. {\bf 27} (1983) 1402; K. Yamagishi,
Nucl. Phys. {\bf B216} (1983) 508; M. Sher, Phys. Repts. {\bf 179}
(1989) 274; C. Ford, D.R.T. Jones, P. -W. Stevenson and M.B. Einhorn,
Nucl. Phys. {\bf B395} (1993) 17 

\bibitem{EO1} E. Elizalde and S.D. Odintsov, Z. Phys. {\bf C64} (1994)
699; Phys. Lett. {\bf B303} (1993) 240
\bibitem{BKMN}M. Bando, T. Kugo, N. Maekawa and H. Nakano,
Prog. Theor. Phys. {\bf 90} (1993) 405
\bibitem{GG1} S.P. Gavrilov and D.M. Gitman,  Phys.
Rev. D {\bf 53} (1996) 7162.
\bibitem{Nik} A.I. Nikishov, Zh. Eksp. Teor. Fiz. {\bf 57}, 1210
 (1969) [Sov. Phys. JETP {\bf 30}, 660 (1970)]; in {\em Quantum Electrodynamics of 
Phenomena
in Intense Fields}, Proc. P.N. Lebedev Phys. Inst. {\bf 111}, 153
 (Nauka, Moscow, 1979)
\bibitem{NaNi} N.B. Narozhny and A.I. Nikishov, Theor. Math. Phys. {\bf
26}, 9 (1976); N.B. Narozhny and A.I. Nikishov, in {\em Problems of
Intense Field Quantum Electrodynamics}, Proc. P.N. Lebedev Phys. Inst. 
{\bf 168}, 175 (Nauka, Moscow 1986)
\bibitem{Fock}V. Fock, Phys. Z. Sowjetunion {\bf 12}, 404 (1937)
\bibitem{GG2} S.P. Gavrilov and D.M. Gitman, 
 J. Math. Phys. {\bf 37}(1996) 3118.

\bibitem{HTF}  {\em Higher Transcendental
functions} (Bateman Manuscript Project),edited by  A. Erdelyi {\em et al.}
 (McGraw-Hill, New York, 1953), Vol. 2.

\bibitem{GMM} A.A. Grib, S.G. Mamaev, and V.M. Mostepanenko, 
{\em Vacuum Quantum Effects in Strong  Fields} (Atomizdat, Moscow
1988; Friedmann Laboratory Publishing, St. Petersburg 1994).

\bibitem{CT} N.A. Chernikov and E.A. Tagirov, Ann. Inst. H. Poincare
{\bf 9A}, 109 (1968)

\bibitem{HH} J.B. Hartle and B.L. Hu, Phys. Rev. {\bf D20} (1979)
1757; M.V. Fischetti, J.B. Hartle and B.L. Hu, Phys. Rev. {\bf D20}
(1979) 1772; P.R. Anderson, Phys. Rev. {\bf D28} (1983) 271; {\bf D29}
(1984) 615

\bibitem{NJL} Y. Nambu and G. Jona-Lasinio, Phys. Rev. {\bf 122}
(1961) 345; W. Bardeen, C. Hill and M. Lindner, Phys. Rev. {\bf D41}
(1990) 1647

\bibitem{KES} Y. Kluger, J.M. Eisenberg, B. Svetitsky, F. Cooper and
E. Mottola, Phys. Rev. {\bf D45}(1992) 4659; {\bf D48} (1993) 190

\bibitem{Par} L. Parker, in {\em Recent Developments in Gravitation},
eds. S. Deser and M. Levy, NY, 1978; S. Fulling, {\em Aspects of
Quantum Field Theory in Curved Space-Time}, (Cambridge Univer. Press,
1989); for earlier work, see L. Parker, Phys. Rev. {\bf 183} (1969) 1057

\bibitem{30} K. Milton, Ann. Phys. (USA) {\bf 127} (1980) 49;
I. Brevik, Ann. Phys. (USA) {\bf 138} (1982) 36; for a review, see
E. Elizalde, S.D. Odintsov, A. Romeo, A.A. Bytsenko and S. Zerbini,
{\em Z\'eta Regularization with Applications}, (World Sci., Singapore,
1994)


\end{thebibliography}
\end{document}